\documentclass[reprint,aps,pre,showkeys] {revtex4-1}
\usepackage{times}
\usepackage{amssymb,amsmath}
\usepackage{graphicx}
\usepackage{mathtools}

\graphicspath{{Plots/}{Schematics/}}

\begin{document}

\title{The structure of infectious disease outbreaks across the animal--human interface}
\author{Sarabjeet Singh}
\email{ss2365@cornell.edu}
\affiliation{Theoretical And Applied Mechanics, Sibley School of Mechanical and Aerospace Engineering, Cornell University}
\author{David J.\ Schneider}
\email{dave.schneider@ars.usda.gov}
\affiliation{Robert W.\ Holley Center for Agriculture and Health,
  Agricultural Research Service, United States Department of
  Agriculture, and Department of Plant Pathology and
  Plant-Microbe Biology, Cornell University, Ithaca, NY 14853}
\author{Christopher R. Myers}
\email{c.myers@cornell.edu}
\affiliation{Computational Biology Service Unit, Institute for Biotechnology and Life Science Technologies, and Laboatory of Atomic and Solid State Physics, Department of Physics, Cornell University}


\begin{abstract}
  Despite the enormous relevance of zoonotic infections to worldwide
  public health, and despite much effort in modeling individual
  zoonoses, a fundamental understanding of the disease dynamics and
  the nature of outbreaks arising in such systems is still lacking. We
  introduce a simple stochastic model of
  susceptible-infected-recovered dynamics in a coupled animal-human
  metapopulation, and solve analytically for several important
  properties of the coupled outbreaks. At early timescales, we solve
  for the probability and time of spillover, and the
  disease prevalence in the animal population at spillover as a
  function of model parameters. At long times, we characterize the
  distribution of outbreak sizes and the critical threshold for a
  large human outbreak, both of which show a strong dependence on the
  basic reproduction number in the animal population. The coupling of
  animal and human infection dynamics has several crucial
  implications, most importantly allowing for the
  possibility of large human outbreaks even when human-to-human
  transmission is subcritical.
  
\end{abstract}



\maketitle

\section{Introduction}
\label{sec:introduction}

Zoonoses -- infectious diseases that spill over from animals to
humans -- represent a major challenge in public
health~\cite{woolhouse2006host,kuiken2005pathogen,lloyd2009epidemic}.
More than half of all known human pathogens are believed to be
zoonotic~\cite{woolhouse2006host}, and zoonotic pathogens are associated
with an overwhelming majority of emerging infectious
diseases~\cite{woolhouse2006host,greger2007human,jones2008global}.
With the confluence of increased disruptions to wildlife
ecosystems and the globalization of human travel, the threat of a
zoonotic pandemic is not only heightened, but is increasingly a part of
the public's consciousness.

Recent research has sought to characterize and classify the
salient features of zoonoses.  Wolfe \emph{et al.}\ proposed a framework to describe evolutionary stages through which  pathogens might evolve from infecting only animals to infecting only humans~\cite{wolfe2007origins}.  Lloyd-Smith \emph{et al.}\ advocated a refinement of that framework emphasizing the importance of the value of the basic reproduction number $R_0$, the average number of new human infections caused by an infectious human host in a fully susceptible human population, in order to distinguish among intermediate stages that transmit to varying degrees in both animals and humans~\cite{lloyd2009epidemic}.  Morse \emph{et al.}\ suggested a different classification that emphasizes the dynamics of infection rather than pathogen properties, distinguishing ``pre-emergence'' (typically spillover from one animal host to another due to changes in habitat or land use) from ``localized emergence'' (transmission into human populations)~\cite{morse2012prediction}.  While these
frameworks are all useful for suggesting further inquiry
(including ours), they are mostly descriptive in nature, and
since they are not tied to specific models of cross-species infection, 
they cannot by themselves be probed in further
quantitative detail.

Mathematical models of zoonotic outbreaks are of increasing
interest, but many important gaps still remain.  The compilation of
Lloyd-Smith \emph{et al.}\ summarized 442 published mathematical models
of various zoonotic diseases, concluding that models that
explicitly incorporate cross-species spillover dynamics are
``dismayingly rare'', despite the fact that such events are the
defining characteristic of zoonotic infection~\cite{lloyd2009epidemic}.  Many of the models summarized that do explicitly include cross-species spillover are risk-based models describing food-borne illness, with fluxes of infection related to some unknown level of initial contamination. These are essentially static, and are thus not applicable in situations where infection prevalance in the animal population is itself dynamic, as would be important for emerging zoonotic diseases. Finally, stochastic treatments of spillover dynamics
are much less common than deterministic models, a fact echoed in a recent survey by Allen \emph{et al.}~\cite{allen2012mathematical}.  Stochastic effects are expected to play dominant roles in outbreak dynamics immediately after an initially unknown number of primary (cross-species) infections have taken place.  During this short period of time, public officials must make critical decisions based on limited and incomplete data.  Truly informed decision making is not possible in the absence of appropriate quantification of uncertainties.

We address these gaps by analyzing a minimal stochastic model of 
directly transmitted zoonoses that explicitly incorporates cross-species
transmission. Our model is restricted to epizootic situations where infection is dynamic in the animal population, as might occur with the introduction of a disease into an amplifier animal host population~\cite{collinge2006disease,childs2007introduction} or with the
emergence of a new, more virulent strain of an existing animal
pathogen~\cite{epstein1995emerging,karesh2012ecology}.  (Thus, the model is currently not applicable to endemic animal diseases that
present an approximately constant force of infection to humans.) 
In these coupled animal-human outbreaks, the degree of human-to-human transmission ($R_0^{hh}$ in our terminology, see below) is no longer the sole determinant of infection prevalance in the human population, but the degree of animal-to-animal transmission and animal-to-human transmission also become important. Formally, our model of zoonoses is an instance of a multitype branching process, and using well-established mathematical techniques, we obtain exact results for many important properties of cross-species outbreaks without needing to examine extensive ensembles of stochastic numerical simulations. While a solution to the full nonlinear problem is not forthcoming, analytical results can be derived in several important limits. At short times, we solve for the distribution of time to spillover into the human population as well as the distribution of the prevalence of animal infections at the time of spillover. Asymptotically at long times where we can characterize the distribution of outbreak sizes in the human population, we identify a parameter regime where large outbreaks are possible in human populations -- \emph{sustained by repeated introductions from the animal population} -- even if human-to-human transmission is subcritical (i.e., when $R_0^{hh}<1$).  Information only about infection in the human population is insufficient to distinguish such a scenario from one involving a single primary introduction followed by extensive human-to-human transmission (see fig. \ref{fig:model_schematic} \textit{bottom}). Our systematic characterization of the spectrum of possible behaviors helps to augment and clarify the previously proposed frameworks.  As with the
classification in \cite{morse2012prediction}, we are ultimately
interested the phenomenology of infection dynamics.  But by tying those 
dynamics explicitly to a mechanistic model for cross-species infection,
we aim to connect that phenomenology to particular regions in the
model's parameter space, such as was advocated in
\cite{lloyd2009epidemic}.  We see our work as a stepping stone toward 
more complex and realistic models that might help others to 
address the spatial and ecological aspects of zoonotic emergence, the evolution of virulence, and public health interventions in the form of dynamic control strategies.

\section{Model}
\label{sec:model}


\begin{figure}[tbp]
\centering
\includegraphics[width=\columnwidth]{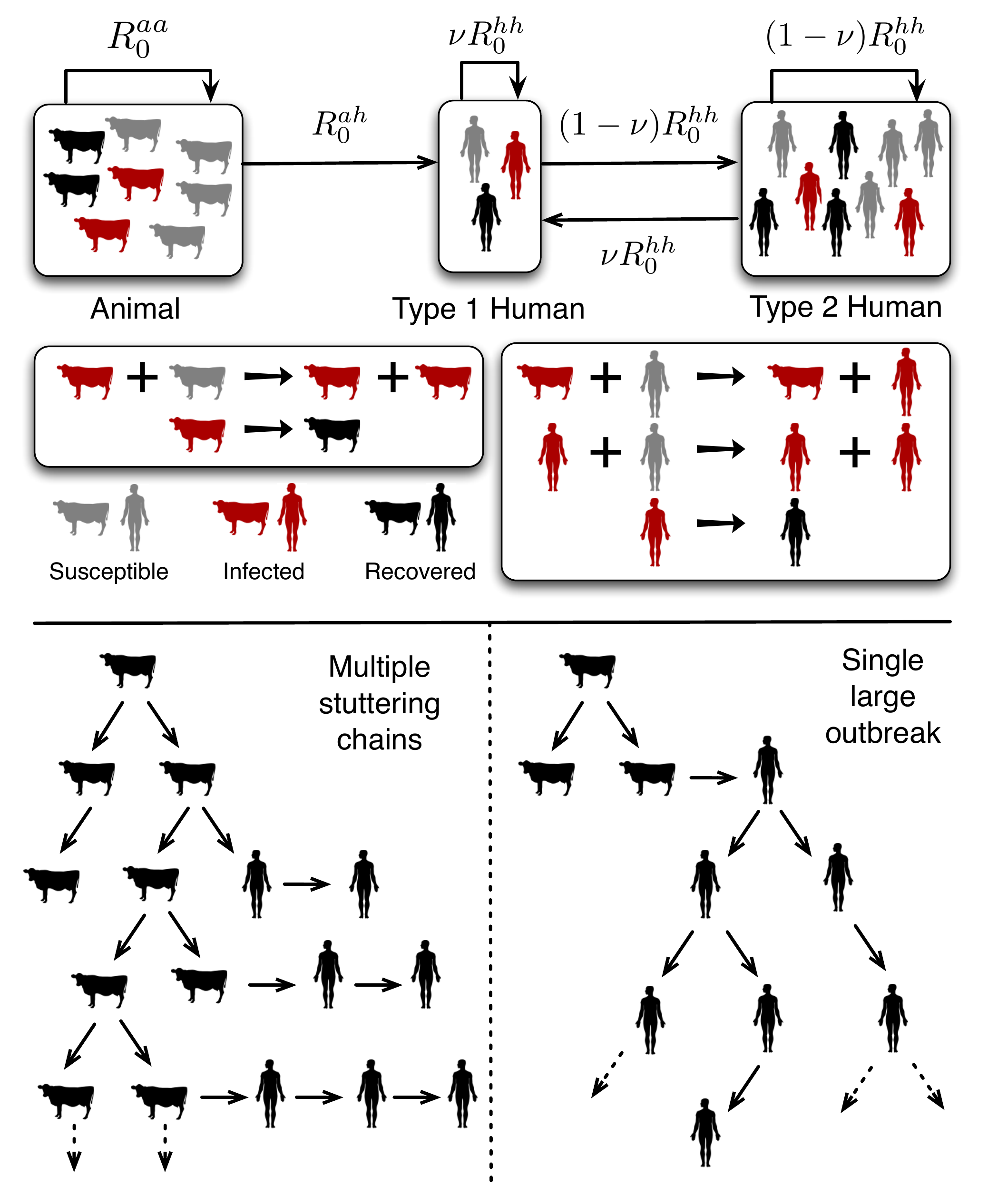}
\vspace{-.1in}
\caption{\label{fig:model_schematic} \textit{Top}: Schematic of our zoonoses model (see Figure \ref{fig:SIRrxn} in \textit{Materials and Methods}). The three--type metapopulation model consists of animals, type 1 humans and type 2 humans. Type 1 humans can receive both primary transmissions from animals and secondary transmissions from other humans, whereas type 2 humans can only receive a secondary transmission. Type 1 humans are fully mixed with both animals and type 2 humans. The arrows denote $R_0$s for inter- and intra-population transmission. \emph{Bottom}: Schematic depicting two possible mechanisms for zoonotic outbreaks in human populations: (Left) Infection spreads efficiently in the animal population but inefficiently in humans, with each introduction into humans leading to a stuttering chain that goes extinct. (Right) An initial spillover leads to a large outbreak sustained by human-to-human transmission.}
\end{figure}


We focus here on spillover into humans from animal hosts on
relatively short timescales, where the prevalence of infection in animals changes quickly relative to other processes, such as 
demographic changes in the host populations, host and pathogen evolution, etc. We assume that the animal hosts are not the natural reservoirs for the pathogen but receive the infection through a rare event, and that infection can be passed on directly to human hosts. The model does not include the possibility of ``spillback'' or reverse infection from humans to animals, an assumption which applies to most zoonoses.

Ours is a multitype stochastic susceptible-infected-recovered
(SIR) model where the two host populations, animal and human, are
fully mixed within their respective species, with a partial overlap
between the species. The partial overlap or the `mixing fraction',
$\nu$, represents the fraction of human hosts that are fully mixed
with the animal hosts (which we denote as `Type 1 humans'). Figure~\ref{fig:model_schematic} (top) shows a schematic of the model and of the underlying SIR reactions; all reactions and reaction rates (probabilities per unit time) are summarized in \textit{Materials and Methods}. The three types of possible infection transmission reactions are animal-to-animal ($aa$), animal-to-human ($ah$), and human-to-human ($hh$).

$R_0^{aa}$ and $R_0^{hh}$ are the basic reproduction numbers (average
number of new infections in a fully susceptible population)
corresponding to within-species infection (eq. \ref{eq:R0_aa_hh} in \textit{Materials and Methods}). $R_0^{ah}$ is defined similarly for the cross-species interaction, i.e., the average number of new infections produced by a single infected animal host in the partially-mixed, fully-susceptible human population (eq. \ref{eq:R0_ah} in \textit{Materials and Methods}). Our results are valid in the limit of large system size for fixed ratio of animal and human population size ($N_a/N_h = \rho$) , where the model reduces to a special case of a multi-type branching process (see \textit{Materials and Methods}). The model includes no explicit time dependence and thus, the time of introduction of infection into the animal population can be taken as $t=0$.  This might occur, for example, following sudden ecological shifts that faciliate a species jump from wildlife to livestock, or the appearance of a novel mutation that provides a mechanism for an endemic pathogen to transmit more effectively in the animal reservoir.

In the absence of cross-species infection, our model would describe two uncoupled SIR processes. The stochastic SIR model has been widely studied \cite{brauer2008mathematical}, and we recount here some of its salient features. An outbreak is defined to be small (or self-limiting) if the total number of hosts infected is $o(N)$ in the limit of infinite system size, $N \rightarrow\infty$, i.e., its relative size does not scale with $N$. Outbreaks are small with probability $1$ below the critical threshold of $R_0 = 1$, whereas above the critical threshold this probability is strictly positive but less than $1$. The corresponding defect in the probability mass is the probability of large outbreaks with characteristic sizes $\mathcal{O}(N)$. The average outbreak size diverges and the distribution of outbreak sizes shows a power-law scaling at the critical threshold.

\section{Results}
\label{ssec:results}

\subsection{Probability of spillover}
A spillover event involves one or more primary
infections in human hosts following the introduction of the disease into
the animal population at $t=0$. The asymptotic ($t\rightarrow\infty$)
probability of spillover as a function of relevant model parameters (eq. \ref{eq:P_spill} in \emph{Appendix}) is shown in blue in Figure~\ref{fig:P_spill}; also shown in gray is the probability of spillover given that there is a small outbreak in the animal population (eq. \ref{eq:P_spill_sm_outb} in \emph{Appendix}). The probability of spillover is less than 1 because the outbreak can die out in the animal population before any primary human infections occur.  Deterministic models associate spillover events with large outbreaks in the animal population. While large outbreaks do enhance the risk of spillover, small outbreaks also contribute. This result indicates that some spillovers may be almost impossible to trace back in the animal population if they arise from a small outbreak where only a few animal hosts were infected and no contact tracing data is available.
\begin{figure}[tbp]
\centering
\includegraphics[width=\columnwidth]{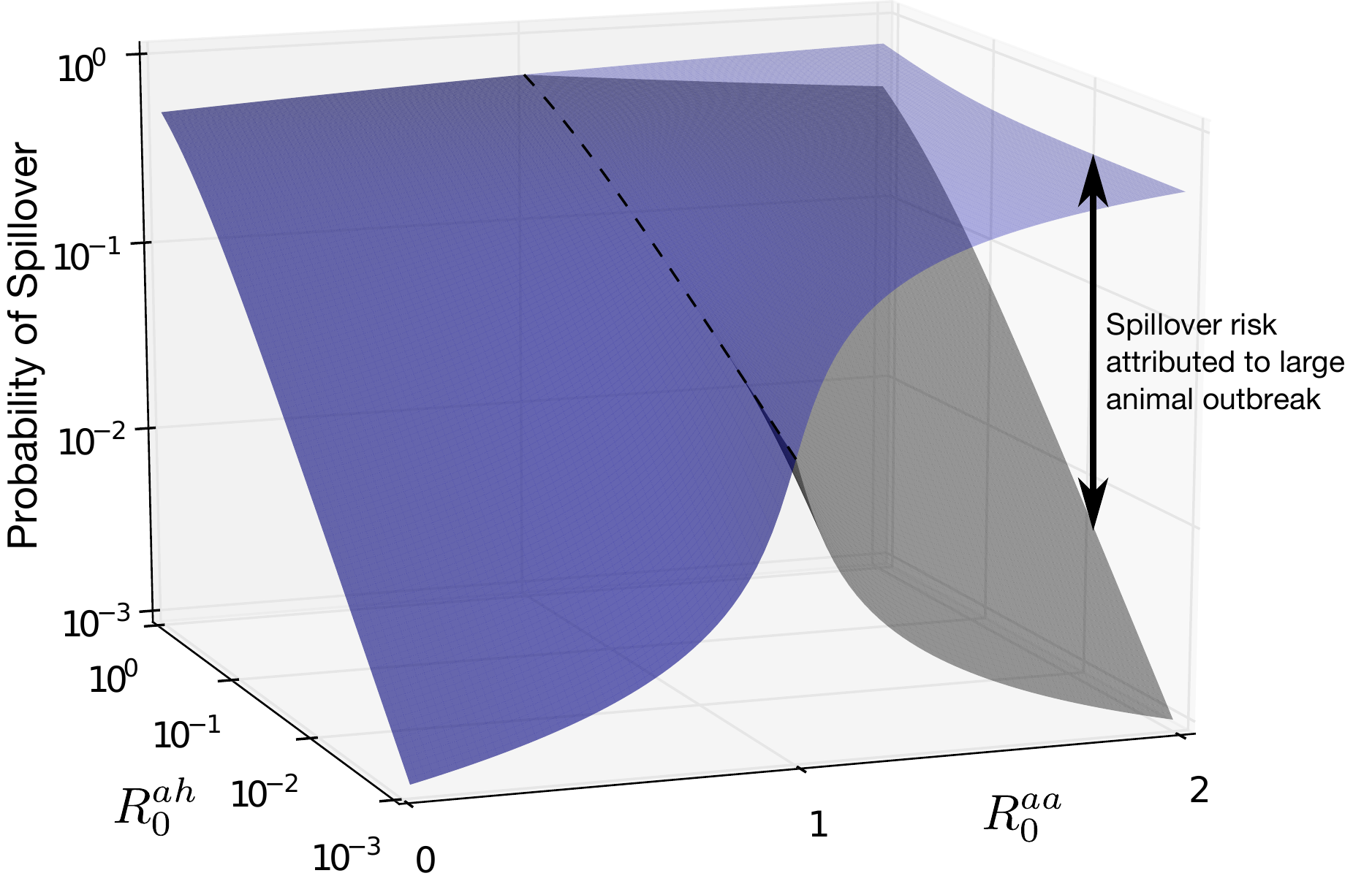}
\vspace{-.15in}
\caption{\label{fig:P_spill} Probability of
  spillover (blue) and the conditional probability of spillover given
  a small outbreak in the animal population (gray). The dashed line marks the separation between the two surfaces at $R_0^{aa} = 1$. The difference between the two surfaces gives the contribution of large animal outbreaks to spillover risk.}
\end{figure}
\begin{figure}[tbp]
\centering
\includegraphics[width=\columnwidth]{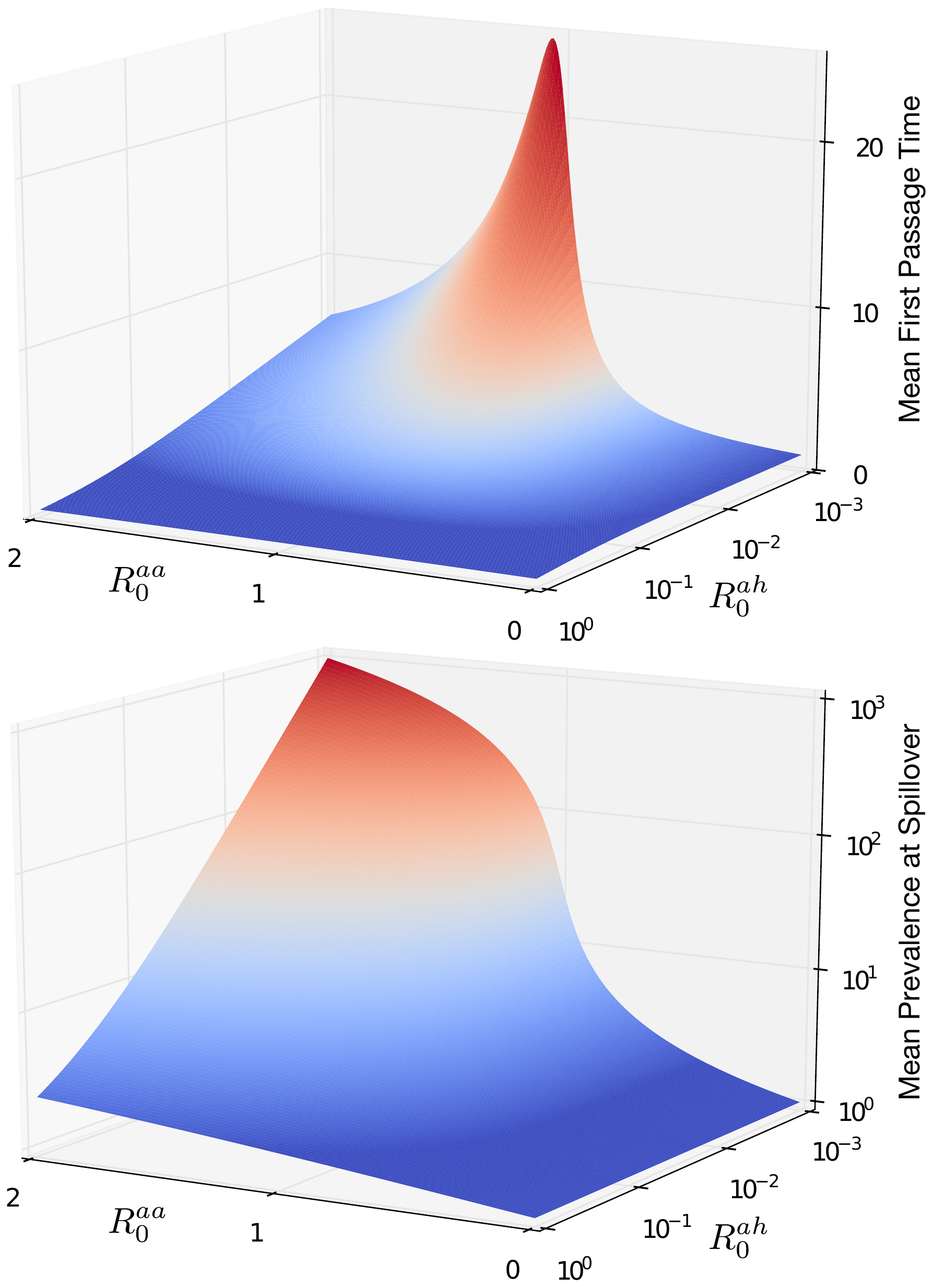}
\vspace{-.15in}
\caption{\label{fig:first_pass_prev_comb} \emph{Top}: The mean time to spillover (in units of the mean infectious period of animal hosts) as a function of $R_0^{aa}$ and $R_0^{ah}$. Coloring represents the standard deviation of the distribution (red:high, blue:low spanning the range $[0.4,18.8]$ on a $\mathrm{log}$ scale). \emph{Bottom}: The number of infectious animal hosts at the time of first primary human infection, with coloring representing the standard deviation of the distribution (red:high, blue:low spanning the range $[10^{-2},10^3]$ on a $\mathrm{log}$ scale).}
\end{figure}
\subsection{Time to spillover (first passage time)} 
In stochastic models, spillover between populations
involves a time delay, the so-called first passage time for spread
into the human population. Figure~\ref{fig:first_pass_prev_comb} (top) shows the mean (surface plot) and standard deviation (colormap) of the first passage time distribution (see \textit{Appendix} for derivation). The distribution is conditional on a spillover taking place, leading to a non-monotonic dependence on $R_0^{aa}$. First passage times for $R_0^{aa} < 1$ are limited by the timescale for the eventual extinction in the animal population: spillover must occur quickly if it is going to happen at all. The expected time to extinction in the animal population diverges as $R_0^{aa} \rightarrow 1$ leading to an increase in the mean first passage time. The mean also decreases with increasing $R_0^{ah}$
because of the increasing rate of animal-to-human transmission. The full
distribution is useful for understanding the relevant timescales
of spillover and their stochastic fluctuations. This serves two
important purposes: first, it indicates whether demography should be factored into the model (i.e., whether spillover will take place on a timescale fast compared to demographic changes), and more crucially, it suggests strategies for optimal surveillance in the field to pinpoint the relevant timescales and surveillance frequencies needed to identify 
emerging zoonotic infections.
\subsection{Disease prevalence in animals at first passage time}
In the absence of animal surveillance, the first spillover into
humans is usually the point at which the disease is first detected and
control interventions are initiated \cite{karesh2012ecology}. While the
first passage time reveals the timescale of spillover, the disease
prevalence reveals the state of the system at spillover. The mean and
the standard deviation of the number of infectious animal hosts at first
passage time $I_a(T)$ are shown in figure~\ref{fig:first_pass_prev_comb}
(bottom). (A similar plot for the number of recovered animal hosts is shown in \textit{Appendix}  figure~\ref{fig:mean_rec_at_spill_log}). Given $I_a(T) = n$, 
maximum-likelihood estimation using this distribution yields a relationship among model parameters: $R_0^{aa} = (n-1)(R_0^{ah} + 1/n)$.
Assuming disease detection coincides with the first spillover event, 
interesting conclusions can be drawn. For $R_0^{aa}$ close to 1, the disease is likely to be detected late, but there will be a low prevalence in the animal population at that time. This is encouraging for public health interventions aimed at controlling the disease in the animal population, although the long delay before detection might 
provide the pathogen sufficient time to evolve greater virulence. For larger $R_0^{aa}$ the spillover is likely to happen relatively early, but the disease prevalence may be quite large, making control
difficult.  Our results indicate that the fluctuations in the prevalence at spillover increase with $R_0^{aa}$, in contrast to the first passage time which has the highest fluctuations near $R_0^{aa} = 1$. This highlights the intrinsic challenges to parameter estimation in order to build predictive models based on prevalence information.

\subsection{Small outbreaks and critical threshold}
\begin{figure}[tbp]
\centering
\includegraphics[width=\columnwidth]{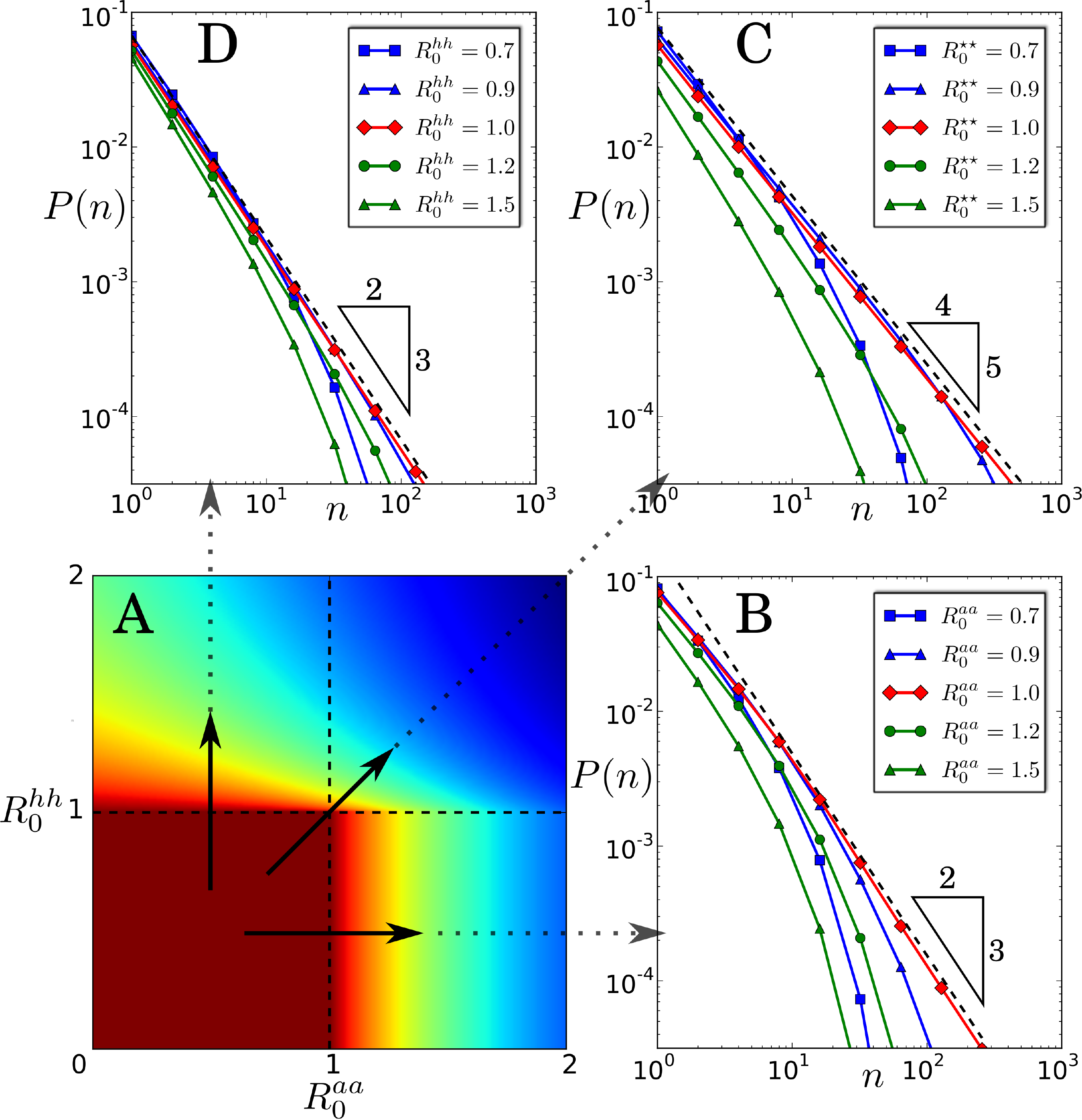}
\vspace{-.15in}
\caption{\label{fig:stut_composite} The distribution of sizes of small
  human outbreaks. (A) Heat map for the probability that an outbreak
  in the human hosts is small spanning the range [0.36 (blue), 1.0 (red)]. (B-D) Probability of having a small outbreak of size $n$ at different crossings of the threshold boundary. All results are for fixed $R_0^{ah} = 0.1$}
\end{figure}

Unlike SIR dynamics in a single population, in our multi-species SIR
model the expected outbreak size diverges if either $R_0^{aa}$ or
$R_0^{hh}$ exceeds 1 (i.e., the threshold is at $\max(R_0^{aa},R_0^{hh}) = 1$; see \textit{Appendix}). Thus, large outbreaks in the human population are possible even if $R_0^{hh} < 1$, which introduces the notion of \emph{spillover-driven large outbreaks}. As in the case of the single-type SIR, small outbreaks occur with nonzero probability throughout the parameter space of our multitype model, albeit with decreasing probability as the system moves beyond the critical threshold. Figure~\ref{fig:stut_composite}A depicts the
probability of a small outbreak plotted against $R_0^{aa}$ and
$R_0^{hh}$ for a fixed $R_0^{ah}$. Also plotted in
Figure~\ref{fig:stut_composite}B-D are the distributions of small
outbreak sizes, which exhibit power-law scaling behavior at the critical
threshold boundary $\mathrm{max}(R_0^{aa},R_0^{hh}) = 1$. There is a
line of critical points at $R_0^{aa} \!=\! 1,\, 0 \!<\!R_0^{hh} \!<\! 1$ and at $R_0^{hh} = 1, 0 < R_0^{aa} < 1$; along both these lines, the scaling behavior is as in a simple SIR model, with the probability of observing an outbreak of size $n$ decaying as $P(n) \sim n^{-3/2}$ (Figure~\ref{fig:stut_composite} B \& D). While identical in the outbreak size scaling, the two lines of the threshold boundary differ on the scaling of the \emph{average} outbreak size which scale as $\mathcal{O} (\min(N_a^{1/3},N_h^{1/2}))$ for $R_0^{aa} \!=\! 1, \, 0 \!<\! R_0^{hh} \!<\! 1$ and $\mathcal{O} (N_h^{1/3})$ for $R_0^{hh} = 1, 0 < R_0^{aa} < 1$. This has implications for determining whether the outbreak is spillover-driven or intrinsically driven: for $N_h \gg N_a^{2/3}$, the abundance of animal hosts, rather than the human hosts, would be a stronger predictor of the human outbreak size. Secondly, for $N_h \ll N_a^{2/3}$, a spillover-driven outbreak has a greater extent of $\mathcal{O}(N_h^{1/2})$ as compared to an intrinsically driven outbreak which is capped at $\mathcal{O}(N_h^{1/3})$. Figure~\ref{fig:stut_composite}C demonstrates that the system exhibits a different scaling behavior, $P(n) \sim n^{-5/4}$, at the multicritical point $R_0^{aa} = R_0^{hh} = 1$ with the average outbreak size scaling with population sizes $N_a$ and $N_h$ as $\mathcal{O}(\min (N_h^{3/4}, (N_a N_h^3 )^{1/7} ))$. See \textit{Appendix} for derivation of these results, as well as comparisons between analytical results and simulations for finite-size systems.  At the multicritical threshold, the outbreak sizes for the epidemics in the animal and human populations diverge simultaneously, resulting in a new universality class with a different scaling behavior.  Our minimal model has washed out most of the small-scale details underlying a zoonotic infection, but we expect -- as is the case with other continuous phase transitions in statistical physics\cite{sethna2006statistical} -- that many of those details will be irrelevant in determining the scaling behavior near the critical threshold. In this regard, we note that same $n^{-5/4}$ scaling -- arising from one critical process driving another -- has been reported recently in a different, albeit related, multitype critical branching process intended to model multistage SIR infections~\cite{antal2012outbreak}.
\subsection{Large human outbreak}
\begin{figure}[tbp]
\centering
\includegraphics[width=\columnwidth]{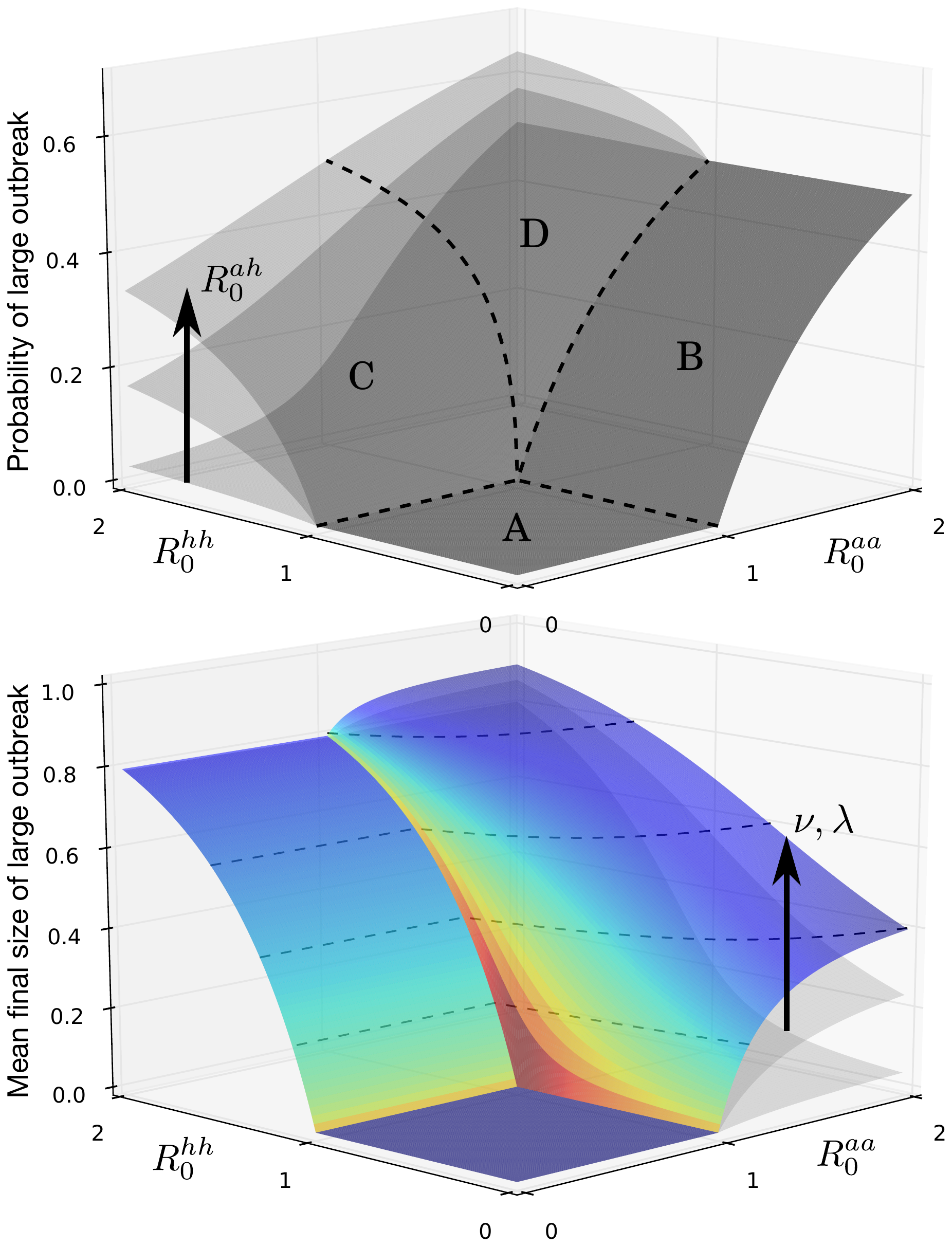}
\vspace{-.15in}
\caption{\label{fig:P_epd_final_size_comb} \emph{Top}: Probability of
  a large outbreak in humans for increasing values of
  $R_0^{ah} = [0.05,0.4,1.0]$. The upper surface is partitioned into 4 sections: (A) where all outbreaks are small, (B) where spillover-driven large outbreaks are possible, (C) where large outbreaks can only be sustained by human to human transmission, and (D) where sustained spillover and human to human transmission result in a large outbreak.
\emph{Bottom}: The mean final size of a large human outbreak plotted against $R_0^{aa}$ and $R_0^{hh}$. The three surfaces are plotted for fixed $\nu = 0.5$ and $\rho = 1$, and increasing $R_0^{ah} = [0.05,0.4,1.0]$. Increasing $\nu$ or $\lambda$ (c.f. eq \ref{eq:lambda}) would result in the same qualitative change in the shape of the surface. Heat map on the upper surface is colored according to the $\mathrm{log}\, \alpha $ (values: [0.5 (blue), 182 (red)]) where $\alpha /\sqrt{N_h}$ is the standard deviation of relative final size in the limit of large $N_h$. The dashed lines on the uppermost surface represent the contours at final size $= [0.2,0.4,0.6,0.8]$}
\end{figure}

Above threshold, there is a nonzero probability of large
outbreaks in the human population. This probability (Eq.~\ref{eq:Q} in
\textit{Materials and Methods}) is shown in
Figure~\ref{fig:P_epd_final_size_comb} (\textit{top}) as a surface
plot in the $R_0^{aa}$ -- $R_0^{hh}$ plane for different values of
$R_0^{ah}$. Region B represents the probability
of a spillover-driven large outbreak. As noted previously, these
outbreaks are `large' despite $R_0^{hh} < 1$, which makes a
classification of zoonotic infection based solely on $R_0^{hh}$
insufficient for this system. As demonstrated in \textit{Appendix}
(figure~\ref{fig:sim_comp}), the dynamics of a large outbreak driven by repeated spillover events can be almost indistinguishable from one dominated by human-to-human transmission. Region C in figure~\ref{fig:P_epd_final_size_comb} (\textit{top}) represents the
probability of a large human outbreak resulting from a finite number $\left(o(N)\right)$ of primary infections but sustained only by
human-to-human transmission ($R_0^{hh} > 1$). The probability shows
dependence on all three $R_0's$, in contrast to the simple SIR where the
probability of large outbreak is simply $1 - 1/R_0^{hh}$. Region D shows 
probability of a large outbreak resulting from the confluence of
repeated spillovers and sustained human-to-human transmission.
The mean fraction of hosts that are infected during a large outbreak
is given by the solution of the transcendental equations for $f_a$
(animals) and $f_h$ (humans):
\begin{subequations}
\begin{align}
	1 - f_a &= e^{-R_0^{aa} f_a} \label{eq:fa} \\
	1 - f_h &= (1 - \nu + \nu e^{-\lambda f_a}) e^{-R_0^{hh} f_h} \label{eq:fh} \\
	\lambda &\equiv \rho R_0^{ah}/\nu \label{eq:lambda}
\end{align}
\end{subequations}
Equation \ref{eq:fa} is the well-known equation for the final size of a single-type SIR. Note that equation \ref{eq:fh}, in the limits of $\nu \rightarrow 0$, $\lambda \to 0$, or $R_0^{aa} \to 0$ reduces to the case a simple SIR as well. All these limits represent an outbreak where the spillover only acts as the conduit for introduction of pathogen in the human hosts but does not affect the dynamics or final size. As can be verified in equation~\ref{eq:fh}, the final size is positive for $R_0^{hh} < 1$, provided $\nu,\lambda,R_0^{aa} > 0$.  These are the necessary conditions for a spillover-driven large outbreak to occur. Interestingly, primary infections directly transmitted from animals may not always dominate the composition of human outbreaks even when $R_0^{hh} < 1$.  Only for $R_0^{hh} < 1/2$ do primary infections occur more frequently than secondary ones on average.  For $1/2 < R_0^{hh} < 1$, there are regions in parameter space where secondary infections could dominate even though the human outbreak is driven by the animal epidemic (see \textit{Appendix} for proof and discussion). For small outbreaks, secondary infections strictly dominate when $R_0^{hh} > 1/2$, a result that has been noted previously (eq. 4 in \cite{blumberg2013inference}). We also calculate the standard deviation of the final size distribution as a function of model parameters using established methods from the theory of metapopulation models (see \textit{Appendix}). Figure~\ref{fig:P_epd_final_size_comb} (\textit{bottom}) shows the surface plot for the mean final size with a colormap that is a function of the standard deviation, for one set of model parameters. As expected, the fluctuations are the largest near the multicritical point and gradually decrease away from it. The plot also shows how the final outbreak size changes as the parameters $\nu$ and $\lambda$ are varied. 

\section{Discussion}
We have presented and analyzed a stochastic model of coupled
infection dynamics in an animal-human metapopulation.  While some of
these results derive from the existing theory of multitype
birth-death processes \cite{bailey1990elements,griffiths1972bivariate,karlin1982linear},
branching processes \cite{antal2012outbreak,athreya1972ney} and
metapopulation models \cite{ball1993final,britton2002epidemics}, other
results are new, and this work represents the first application of such results to the study of zoonoses. We have described spillover
from animal to human populations, but such a model -- or a variant of
it -- would be applicable to other cross-species infections, such as among different animal hosts. In metapopulation models, the specific form of the inter-population coupling arises from the particular processes or population structure that one aims to address with such coupling. In our model, the existence of a smaller, at-risk population of animal-exposed humans is motivated in particular by the ecology of
animal-human interactions.  Different forms of coupling might be more
applicable to other cross-species infections.

The coupling of animal and human infectious disease dynamics results in important changes to the structure of outbreaks in human populations as compared to those in a human-only SIR model. In the subcritical regime where stuttering chains of transmission dominate, this coupling enhances the probability of longer chains (Figure~\ref{fig:stut_composite}C), which could allow for greater opportunity for pathogen adaptation to human hosts~\cite{lloyd2009epidemic,antia2003role}. The picture that emerges from our analysis of the coupled system is somewhat 
qualitatively different than the zoonotic classification schemes that
have been previously proposed~\cite{wolfe2007origins,lloyd2009epidemic}.  
In particular, a large outbreak can not be attributed solely to $R_0^{hh} > 1$: Stages II, III and IV discussed in ~\cite{wolfe2007origins,lloyd2009epidemic} can all support large outbreaks in the human population if driven sufficiently hard by an animal outbreak. This could have important ramifications for
zoonotic diseases where human-to-human transmission is not the crucial
determinant of the epidemic outcome such as rabies, Nipah, Hendra and
Menangle \cite{karesh2012ecology,morse2012prediction}. The cross-species
coupling also complicates the problem of inference and parameter estimation in the face of a new outbreak. 

In addition, our analysis suggests the need to be precise with other terminology.  The term `stuttering chain' has been used in literature
\cite{lloyd2009epidemic,antia2003role} to describe a chain of
infections starting from a single infectious host that goes extinct
without affecting a significant fraction of the host population. For
the single-type SIR model, the term is synonymous with `small
outbreak' as we have defined here, and the epidemic threshold is the
point in parameter space at which the average length of one such chain
diverges. But in our multitype SIR model, the term `stuttering chain' can not be used interchangeably with `small outbreak'.  Since multiple
introductions can occur in the human population, an outbreak is small
if and only if (1) a finite number of distinct infection chains occur
in the human population, and (2) all such chains stutter to
extinction. A large outbreak in the human hosts occurs when any one of
these conditions is violated. Specifically, a spillover-driven large
outbreak occurs when the number of infection chains diverges, which can
happen if $R_0^{aa} > 1$. Separately, the length of any one such chain
can diverge if $R_0^{hh} > 1$.

The mechanistic details of our model allow us to capture the
phenomenology associated with a wide range of zoonoses.
Transmission rates reflect a number of ecological and 
immunological factors, which can be difficult to disentangle. We have
taken a first step in doing so by explicitly accounting for an at-risk
human subpopulation through the `mixing fraction' $\nu$, which 
describes the geographical or ecological overlap between humans and
animals.  This overlap can vary significantly in different 
situations (e.g., bush-meat trade in sub-Saharan Africa versus poultry and pig farming practices in south-east Asia). The remaining details of animal-human transmission remain embedded in the parameter $R_0^{ah}$, which must be unraveled for any particular disease through further research. Interestingly, the basic reproduction numbers are sufficient to describe the properties of an outbreak on short timescales (e.g., time to spillover, the probability of spillover, and distribution of sizes of small outbreaks), whereas the other model parameters are relevant in determining the attributes of the epidemic
process at longer times (e.g., mean final size and variance of large
outbreaks).

The community has advocated `model-guided fieldwork' \cite{lloyd2009epidemic,restif2012model,wood2012framework}, 
as well as increased collaboration between public health scientists and ecologists in developing integrated approaches to predicting and preventing zoonotic epidemics\cite{karesh2012ecology}.  Mathematical analysis needs to play a central role in such activities, in order to assess the implications of model assumptions.  The model we have analyzed certainly does not describe all of the complexity of cross-species infection, and any more comprehensive theory would need to account for other factors such as the ecology of interactions between wildlife and domesticated animals, the encroachment of human development into animal habitats, the evolution of virulence, and the propensity for pathogens to successfully jump across species.  But in distilling some essential features of cross-species outbreaks, we hope to identify key
aspects of phenomenology, highlight the role of important  
processes, and suggest further inquiry into particular systems
of interest.

\section{Materials and methods}
\label{sec:materials}
\subsection{Model equations}
\begin{figure}[htp]
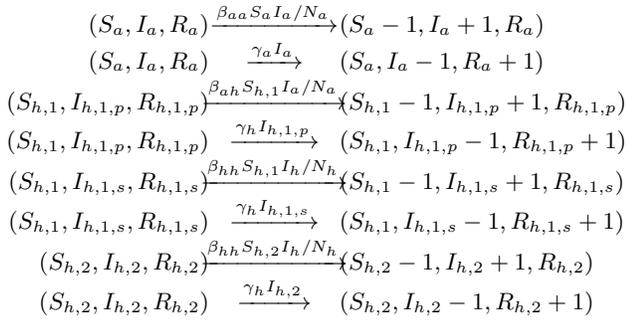
 
\small{
\begin{displaymath} 
{\begin{array}{rcl}
(S_a,I_a,R_a) \!\!\!\!\!\!&\xrightarrow{\beta_{aa} S_a I_a / N_a} &\!\!\!\!\!\!(S_a-1,I_a+1,R_a) \\
(S_a,I_a,R_a) \!\!\!\!\!\!&\xrightarrow{\gamma_a I_a} &\!\!\!\!\!\!(S_a,I_a-1,R_a+1) \\
(S_{h,1},I_{h,1,p},R_{h,1,p})\!\!\!\!\!\!&\xrightarrow{\beta_{ah} S_{h,1} I_a / N_a}&\!\!\!\!\!\!(S_{h,1}-1,I_{h,1,p}+1,R_{h,1,p}) \\
(S_{h,1},I_{h,1,p},R_{h,1,p}) \!\!\!\!\!\!&\xrightarrow{\gamma_h I_{h,1,p}} &\!\!\!\!\!\!(S_{h,1},I_{h,1,p}-1,R_{h,1,p}+1) \\
(S_{h,1},I_{h,1,s},R_{h,1,s}) \!\!\!\!\!\!&\xrightarrow{\beta_{hh} S_{h,1} I_h/N_h} &\!\!\!\!\!\!(S_{h,1}-1,I_{h,1,s}+1,R_{h,1,s})\\
(S_{h,1},I_{h,1,s},R_{h,1,s}) \!\!\!\!\!\!&\xrightarrow{\gamma_h I_{h,1,s}} &\!\!\!\!\!\!(S_{h,1},I_{h,1,s}-1,R_{h,1,s}+1) \\
(S_{h,2},I_{h,2},R_{h,2}) \!\!\!\!\!\!&\xrightarrow{\beta_{hh} S_{h,2} I_h/N_h} &\!\!\!\!\!\!(S_{h,2}-1,I_{h,2}+1,R_{h,2})\\
(S_{h,2},I_{h,2},R_{h,2}) \!\!\!\!\!\!&\xrightarrow{\gamma_h I_{h,2}} &\!\!\!\!\!\!(S_{h,2},I_{h,2}-1,R_{h,2}+1)
\end{array}}
\end{displaymath}}
\vspace{-.075in}
{\caption{ \label{fig:SIRrxn} Model Reactions with rates (probabilities
    per unit time). The subscripts `1' and `2' identify the type of
    human host. Type 1 humans can receive infection from both
    species. Type 2 can only receive infection from humans. The
    subscripts `\textit{p}' and `\textit{s}' distinguish between primary
    and secondary infections, e.g., $I_{h,1,s}$ is the number of
    infected human hosts in type 1 infected via secondary transmission. $I_h$ is the total number of human infections, i.e, $I_h = I_{h,1,p} + I_{h,1,s} + I_{h,2}$.}}
\end{figure}

See figure \ref{fig:SIRrxn} for model reaction equations.

\subsection{Basic reproduction numbers}
\begin{subequations}
\begin{align} 
	R_0^{aa} &= \dfrac{\beta_{aa}}{\gamma_a} , \quad R_0^{hh} = \dfrac{\beta_{hh}}{\gamma_h} \label{eq:R0_aa_hh} \\
	\quad R_0^{ah} &= \dfrac{\nu \beta_{ah}}{\rho \gamma_a} \equiv \dfrac{\hat{\beta}_{ah}}{\gamma_a} \equiv \dfrac{\nu \lambda}{\rho} \label{eq:R0_ah}
\end{align}
\end{subequations}

\subsection{Multitype linear birth-death process}
In the limit of large system size, a subset of the linearized process can be summarized by the following reactions
\small{
\begin{alignat}{3}
	&(I_a,R_a) &\xrightarrow{\beta_{aa} I_a } &(I_a+1,R_a) \notag\\[-1pt]
	&(I_a,R_a) &\xrightarrow{\gamma_a I_a} &(I_a-1,R_a+1) \\[-1pt]
	&(I_h,Z_{h,p},Z_{h,s}) &\xrightarrow{\hat{\beta}_{ah} I_a} &(I_h+1,Z_{h,p}+1,Z_{h,s}) \notag
\end{alignat}}
where $Z_\star = I_\star + R_\star$ in above process. The joint distribution of the process $(I_a,R_a,Z_{h,p})$ is generated from the following PGF (probability generating functions) which has an explicit analytical solution \cite{athreya1972ney,bailey1990elements}. 
\begin{equation*}
	G_{ah}(x,y,z;t) =\!\! \sum \limits_{l,m,n} \! \mathbb{P}[I_a(t) \!=\! l,R_a(t) \!=\! m,Z_{h,p}(t) \!=\! n] \, x^l y^m z^n \\[-5pt]
\end{equation*}
Once the PGF $G_{ah}(x,y,z;t)$ is solved analytically (see \emph{Appendix}), the distribution of primary infections $Z_{h,p}(t)$ can be generated by $G_{ah}(1,1,z;t)$ and that of first passage time $T$ is extracted by noting that 
\begin{equation*}
	\mathbb{P}[T \le t] = \mathbb{P}[Z_{h,p}(t) > 0]
\end{equation*}
The probability of spillover is simply $\mathbb{P}[T < \infty] = \mathbb{P}[Z_{h,p}(\infty) > 0]$. Finite size corrections to the probability of spillover can be calculated analytically. See \emph{Appendix} for details.

\subsection{Multitype branching processes}
The joint distribution of the number of infected animals and primary human infections at the end of an outbreak is generated by $G_{ah}(1,y,z;\infty)$. This yields the following PGFs for the marginal distributions.
\begin{align}
	H_a(y) &= G_{ah}(1,y,1;\infty) \notag\\
	H_{h,p}(z) &= G_{ah}(1,1,z;\infty)
\end{align}
The distributions of primary and secondary infections can be combined assuming tree-like structure for the composite infection chains which gives the nested PGF for the human outbreak sizes.
\begin{equation}
H_h(x) = H_{h,p}(x \hat{H}_{h,s}(x)) \\[-5pt]
\end{equation}
where $\hat{H}_{h,s}(x)$ is the PGF for the secondary transmissions originating from a single primary (\ref{eq:H_hs_hat} in \emph{Appendix}). The probability of a large human outbreak is the defective probability mass in the distribution.
\begin{equation} \label{eq:Q}
	\mathbb{P}[\text{large outbreak}] = 1 - H_h(1)
\end{equation}
\subsection{Finite size effects}
Although the theory is derived in the limit of infinite populations, we find a good agreement with simulations done for system sizes as low as $N_a, N_h = 10^3$. The simulations are done using Gillespie's direct method \cite{gillespie1977exact,keelingmodeling}.


\begin{acknowledgments}
The authors would like to thank Jason Hindes, Oleg Kogan, Marshall
Hayes, Drew Dolgert and Jamie Lloyd-Smith for helpful discussions and comments on the manuscript.  This work was supported by the Science \& Technology Directorate, Department of Homeland Security via interagency agreement no.\ HSHQDC-10-X-00138.
\end{acknowledgments}





\renewcommand{\thefigure}{A\arabic{figure}}
\setcounter{figure}{0}

\appendix

\section{Dynamics of the multi-type birth and death process}
We investigate the multitype SIR model (figure \ref{fig:SIRrxn} in \textit{Materials and Methods}) in the limit of $N_a,N_h \rightarrow \infty, N_a/N_h \rightarrow \rho$. In this limit, the process reduces to a multitype linear birth-death process. Let $Z_{\star}(t) = I_{\star}(t) + R_{\star}(t)$ where $\star$ stands for particular subscripts used in what follows.  $Z_{h,p}(t)$ denotes the number of primary human infections and $Z_{h,s}(t)$ denotes the number of secondary human infections irrespective of the human host type. The total number of infected human hosts is then $Z_h(t) = Z_{h,p}(t) + Z_{h,s}(t)$. The multitype linear birth-death process is summarized by the following reactions.
\begin{alignat}{3}
	&(I_a,R_a) &\xrightarrow{\beta_{aa} I_a } &(I_a+1,R_a) \notag\\
	&(I_a,R_a) &\xrightarrow{\gamma_a I_a} &(I_a-1,R_a+1) \notag\\
	&(I_h,Z_{h,p},Z_{h,s}) &\xrightarrow{\hat{\beta}_{ah} I_a} &(I_h+1,Z_{h,p}+1,Z_{h,s}) \\
	&(I_h,Z_{h,p},Z_{h,s}) &\xrightarrow{\beta_{hh} I_h} &(I_h+1,Z_{h,p},Z_{h,s}+1) \notag\\
	&(I_h,Z_{h,p},Z_{h,s}) &\xrightarrow{\gamma_h I_h}   &(I_h-1,Z_{h,p},Z_{h,s}) \notag
\end{alignat}
where $\hat{\beta}_{ah} \equiv \nu \beta_{ah} / \rho$. The basic reproduction numbers associated with the \emph{aa}, \emph{ah} and \emph{hh} transmissions are (cf. eq. \ref{eq:R0_aa_hh}, \ref{eq:R0_ah} in \emph{materials and methods})
\begin{equation} \label{eq:SI_R0s}
	R_0^{aa} = \dfrac{\beta_{aa}}{\gamma_a} , \quad R_0^{hh} = \dfrac{\beta_{hh}}{\gamma_h}, \quad R_0^{ah} = \dfrac{\nu \beta_{ah}}{\rho \gamma_a} \equiv \dfrac{\hat{\beta}_{ah}}{\gamma_a}
\end{equation}
In our model, we have used the population of the animal hosts to dilute the per-contact rate of transmission, i.e.,
\begin{equation}\label{eq:rate1}
	rate_{\text{A-H}} = \dfrac{\beta_{ah} S_{h,1} I_a}{N_a}
\end{equation}
 Alternatively, one might construct a different version of the model that uses the population of at-risk human hosts to dilute the transmission rate, 
\begin{equation}
	\tilde rate_{\text{A-H}} = \dfrac{\beta_{ah} S_{h,1} I_a}{\nu N_h}
\end{equation}
The choice of a particular cross-species rate depends on the context and the animal-human ecology for a specific disease. We shall proceed with the first description (eq. \ref{eq:rate1}), but note that the results are independent of the choice, as long as the parameter $R_0^{ah}$ is rescaled accordingly.

The distribution of the process can be solved using probability generating functions (PGFs)  \cite{SI_athreya1972ney,SI_bailey1990elements}. Let $G_a(x,y,u,z,w;t)$ be the PGF for the joint distribution of the dynamic variables when a single animal host was infected at time 0. Similarly, let $G_h(u,w;t)$ be the PGF for the joint distribution of $(I_h(t),Z_{h,s}(t))$ where a single human host is infected at time 0 and there is no cross-species transmission.
From \cite{SI_athreya1972ney}, we can write down the following backward equation for these generating functions.
\begin{align}
	\dfrac{\partial G_a}{\partial t} &= U_a(G_a,y,G_h,z,w) \\
	\dfrac{\partial G_h}{\partial t} &= U_h(G_h,w) \notag
\end{align}
where $U_a(x,y,u,z,w)$ and $U_h(u,w)$ are given by
\begin{align}
	U_a(x,\!y,\!u,\!z,\!w) &= \beta_{aa} x^2 + \gamma_a y  - (\beta_{aa} \!+\! \hat{\beta}_{ah} \!+\! \gamma_a) x  + \hat{\beta}_{ah} x u z \notag\\
	U_h(u,w) &= \beta_{hh} u^2 w + \gamma_h - (\beta_{hh} + \gamma_h) u
\end{align}
The initial conditions for this set of equations are
\begin{align}
 	G_a(x,y,u,z,w;0) &= x \notag\\
 	G_h(u,w;0) &= u
\end{align}
The equation for $G_h$ can be solved exactly. The solution is provided in \cite{SI_athreya1972ney,SI_bailey1990elements} and we reproduce it here.
\begin{align} \label{eq:G_h}
	G_h(u,w;t)\! = \!\dfrac{A_h(\!B_h\! -\! u) \!+\! B_h(u \!-\! A_h)e^{-\beta_{hh} w (\!B_h-A_h\!) t}} {(B_h\!-\!u)\! +\! (u\! -\! A_h) e^{-\beta_{hh} w (\!B_h\!-\!A_h\!) t}}
\end{align}
where $A_h(w)$ and $B_h(w)$ are solutions of the following quadratic equation such that $0 < A_h < 1 < B_h$.
\[ R_0^{hh} w s^2 - (R_0^{hh} + 1) s + 1 = 0\]
The PGF $G_h$ quantifies the distribution of a single small outbreak or stuttering chain \cite{SI_lloyd2009epidemic} which is disentangled from the A-H transmission dynamics. The more interesting aspect of the zoonoses dynamics is captured by the first equation (for $G_a$). While a full analytical solution to the process has recently been solved \cite{SI_antal2011exact}, we require the solution to a subset of the complete process as described in the next section.

\subsection{Distribution of primary human infections}
The distribution of $(I_a,R_a,Z_{h,p})$ is governed by a reduced set of reaction equations.
\begin{alignat}{3}
	&(I_a,R_a,Z_{h,p}) &\xrightarrow{\beta_{aa} I_a } &(I_a+1,R_a,Z_{h,p}) \notag\\
	&(I_a,R_a,Z_{h,p}) &\xrightarrow{\gamma_a I_a} &(I_a-1,R_a+1,Z_{h,p}) \\
	&(I_a,R_a,Z_{h,p}) &\xrightarrow{\hat{\beta}_{ah} I_a} &(I_a,R_a,Z_{h,p}+1) \notag
\end{alignat}
Let $G_{ah}(x,y,z;t)$ represent the PGF for the distribution of the above process. Following the methods outlined in \cite{SI_bailey1990elements}, we obtain the following solution to the system. The distribution reported here has been solved before in the context of a human-only epidemic process with two types of hosts \cite{SI_griffiths1972bivariate}.
\begin{align} \label{eq:G_ah}
	G_{ah}(x,\!y,\!z;\!t)\! = \dfrac{A_a(\!B_a \! \!-\! x)\! +\! B_a(x \!-\! A_a)e^{-\beta_{aa} (B_a\!-\!A_a) t}} {(B_a\!-\!x) + (x\! -\! A_a) e^{-\beta_{aa} (B_a\!-\!A_a) t}}
\end{align}
where $A_a(y,z)$ and $B_a(y,z)$ are roots of the following quadratic equation such that $0 < A_a < 1 < B_a$.
\begin{equation} \label{eq:quad}
R_0^{aa} s^2 - \left(R_0^{aa} \!+\! 1 \!+\! R_0^{ah}(1 - z) \right) s + y = 0
\end{equation}
In subsequent sections, we shall require the value of roots at the point $z=0$. Adopting notation from \cite{SI_karlin1982linear}, we define
\begin{align} \label{eq:roots}
	V_0(y) &= A_a(y,0) \quad v_0 = A_a(1,0) \notag\\
	V_1(y) &= B_a(y,0) \quad v_1 = B_a(1,0)
\end{align}

\subsection{First passage time} \label{subsec:first_pass}
\begin{figure} [tbp]
\centering
\includegraphics[width=\columnwidth]{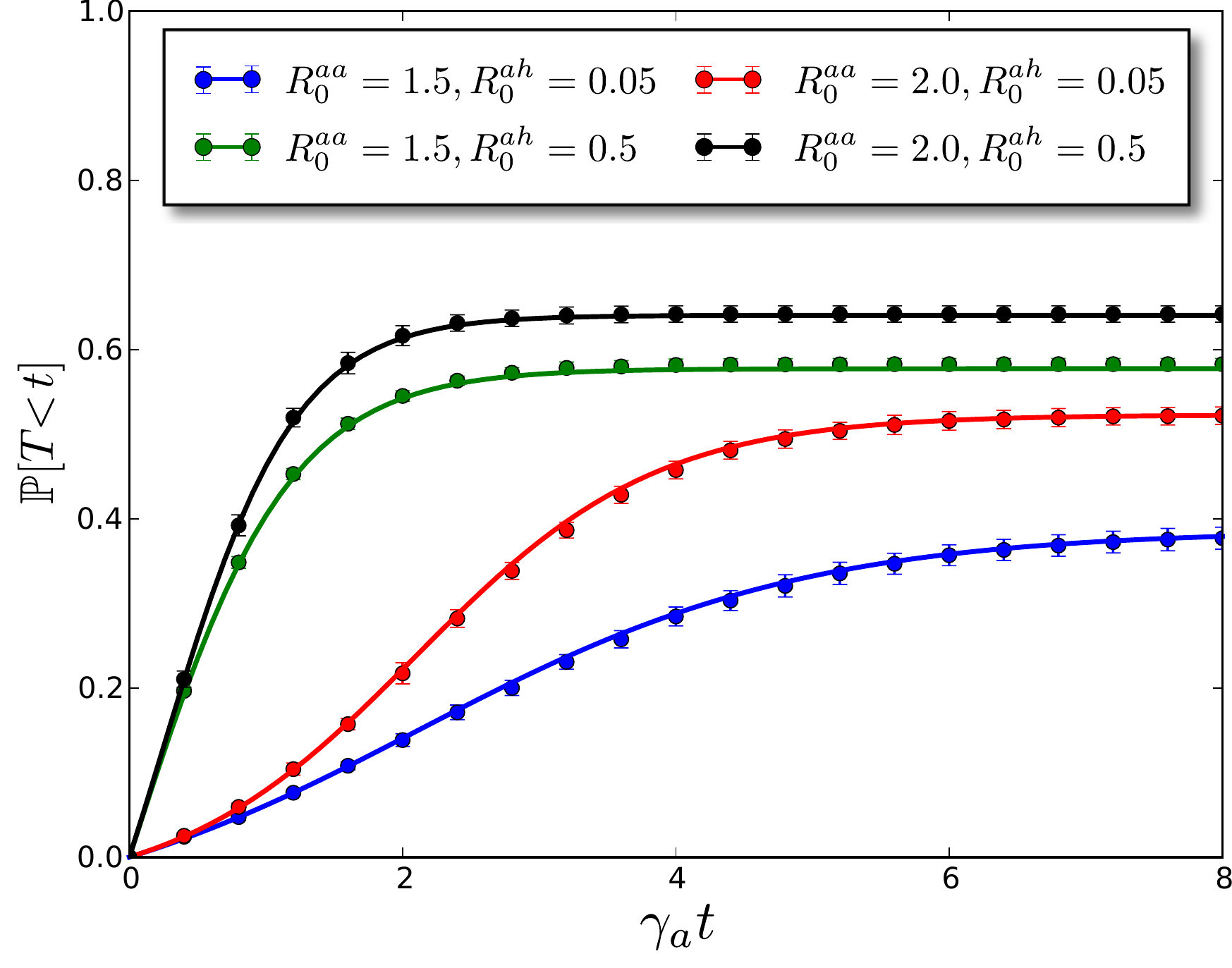}
\caption{\label{fig:sim_vs_ana} Comparison of analytical distribution given by eq. \ref{eq:first_pass_dist} (solid line) with discrete event simulation (Gillespie's direct method) for $\mathbb{P}[T < t]$ with finite system size ($N_a = N_h = 10^3$). X-axis is time normalized by the mean infectious period ($1/\gamma_a$), of the animal species. The markers represent the mean of 8000 simulation runs.}
\end{figure}

\begin{figure} [tbp]
\centering
\includegraphics[width=\columnwidth]{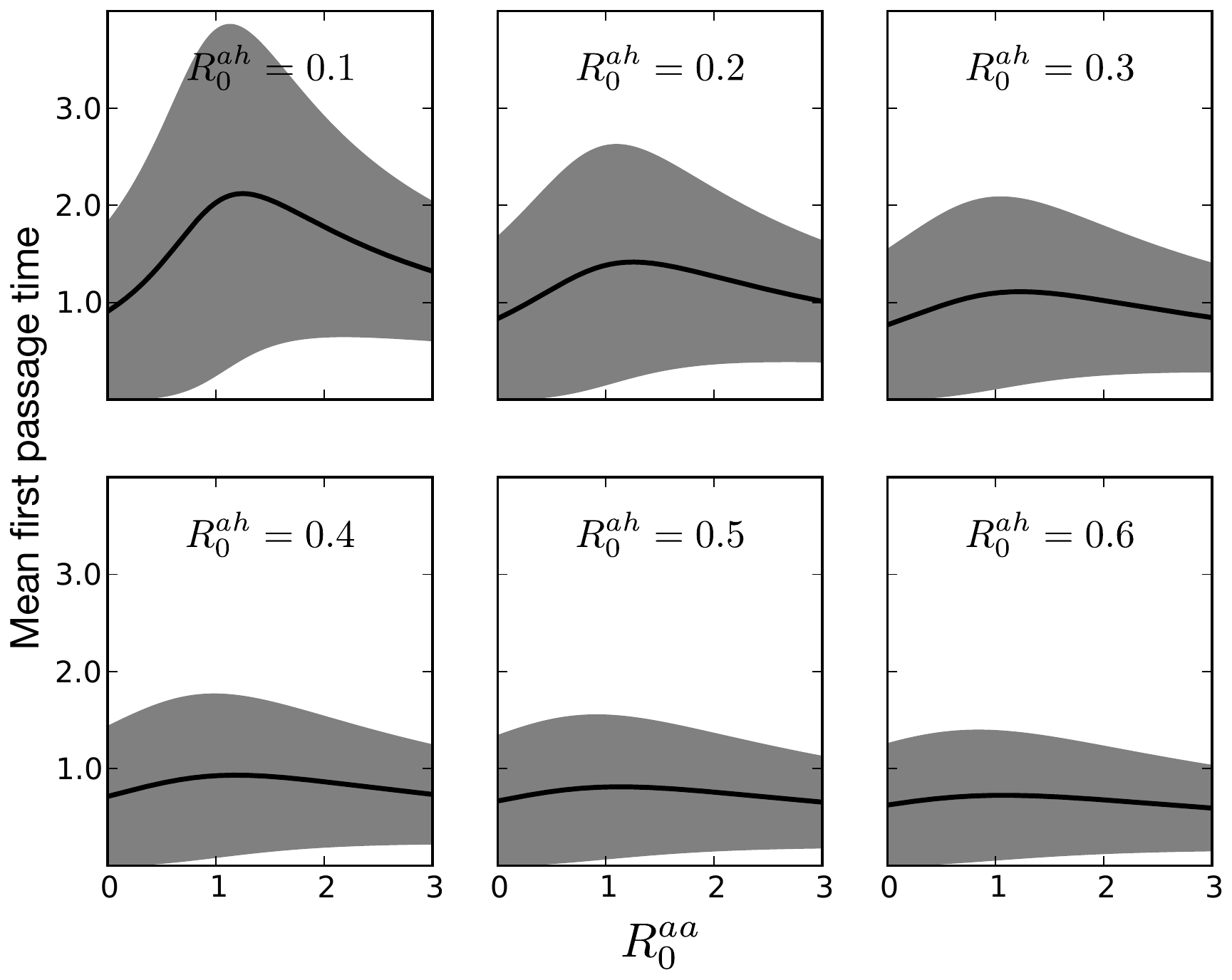}
\caption{\label{fig:first_passage_slice} Mean first passage time plotted against $R_0^{aa}$ for different slices of $R_0^{ah}$. The spread around the mean is one standard deviation of the distribution.}
\end{figure}

We define the time to spillover as the first passage time $T$ for human infection, i.e., 
as the time when the first primary infection occurs in the human hosts.
\begin{align} \label{eq:first_pass_dist}
\mathbb{P}[T \le t]    &= \mathbb{P}[Z_{h,p}(t) > 0] \notag\\
					   &= 1 - 	G_{ah}(1,1,0;t) \notag\\
					   = 1 - &\dfrac{v_0(v_1 \!-\! 1)\! +\! v_1(1\! -\! v_0)e^{-\beta_{aa} (v_1\!-\!v_0) t}} {(v_1-1) + (1 - v_0) e^{-\beta_{aa} (v_1-v_0) t}}
\end{align}	
The distribution is plotted in figure~\ref{fig:sim_vs_ana} along with results of discrete event simulation drawn from the underlying set of reactions. Simulations were done using Gillespie's direct method \cite{SI_keelingmodeling} for reaction kinetics. Figure~\ref{fig:first_passage_slice} shows slices of the mean first passage time surface (figure~\ref{fig:first_pass_prev_comb} in main text) with one standard deviation spread.
It can be seen that the distribution is defective since the disease can go extinct in the animal population before the first primary transmission occurs in the human population. Thus, we can calculate the probability of spillover as
\begin{equation} \label{eq:P_spill}
\mathbb{P}[T < \infty] = 1 - v_0 
\end{equation}
The conditional distribution $\mathbb{P}[T < t \mid T < \infty]$ is
\begin{align} \label{eq:P_spill_cond}
	\mathbb{P}[T < t \mid T < \infty] = \dfrac{1 - e^{-\beta_{aa} (v_1-v_0) t}}{1 \!+\! \left(\dfrac{1 \!-\! v_0}{v_1 \!-\! 1}\right)\! e^{-\beta_{aa} (v_1-v_0) t}}
\end{align}

\subsubsection{Moments of first passage time} \label{sec:first passage time}
\begin{align}
\noindent\mathbb{E}[T^n \!\mid\! T \!<\! \infty] &= \dfrac{\mathbb{E}[T^n \mathbf{1}_{\{T < \infty\}}]}{\mathbb{P}[ T < \infty]} \notag\\
	&= \dfrac{\mathbb{E}[ \left( T \mathbf{1}_{\{T < \infty\}} \right)^n ]}{\mathbb{P}[ T < \infty]} \notag\\
	&= \dfrac{n \int_0^\infty t^{n-1} \; \mathbb{P}[T \mathbf{1}_{\{T < \infty\}} > t] \, dt }{\mathbb{P}[ T < \infty]} \notag\\
	= n (v_1 \!-\! v_0) \!\! \int_0^\infty \!\!\!\!\!& \dfrac{t^{n-1} e^{-\beta_{aa} (v_1-v_0) t} }{(v_1\!-\!1) \!+\! (1 \!-\! v_0) e^{-\beta_{aa} (v_1\!-\!v_0) t}} dt \notag
\end{align}
Let $c = v_1 - 1,	d = 1 - v_0$ and $k = \beta_{aa} (v_1 - v_0)$.
\begin{align}
	\mathbb{E}[T^n \mid T < \infty] &= \dfrac{nk}{\beta_{aa}} \int_0^\infty \dfrac{t^{n-1} e^{-kt}}{c + d e^{-kt}} \; dt
\end{align}
This is the integral of the Bose-Einstein distribution which can be expressed using the polylogarithm function.
\begin{align}	
	&= \dfrac{-n!}{\beta_{aa}  k^{n-1} d} \: \mathrm{Li}_n \left( \dfrac{-d}{c} \right) \notag\\
	&= \dfrac{n!}{(\beta_{aa})^n \left(v_0 - 1\right)\left(v_1 - v_0 \right)^{n-1}} \mathrm{Li}_n \left( \dfrac{v_0 - 1} {v_1 - 1} \right)
\end{align}
where $\mathrm{Li}_n(z)$ is the polylogarithm function of order $n$. Putting $n=1$ in the expression, we obtain the conditional expected value of the first passage time.
\begin{equation}
\mathbb{E}[T \mid T < \infty] = \dfrac{1}{\beta_{aa} \left(1 - v_0\right)} \mathrm{log} \left( \dfrac{v_1 - v_0} {v_1 - 1} \right)
\end{equation}
To our knowledge, only the first moment has been reported earlier in \cite{SI_karlin1982linear}, which was in the context of population genetics.

\subsection{Finite-size corrections to probability of spillover}
\begin{figure}[tbp]
\centering
\includegraphics[width=\columnwidth]{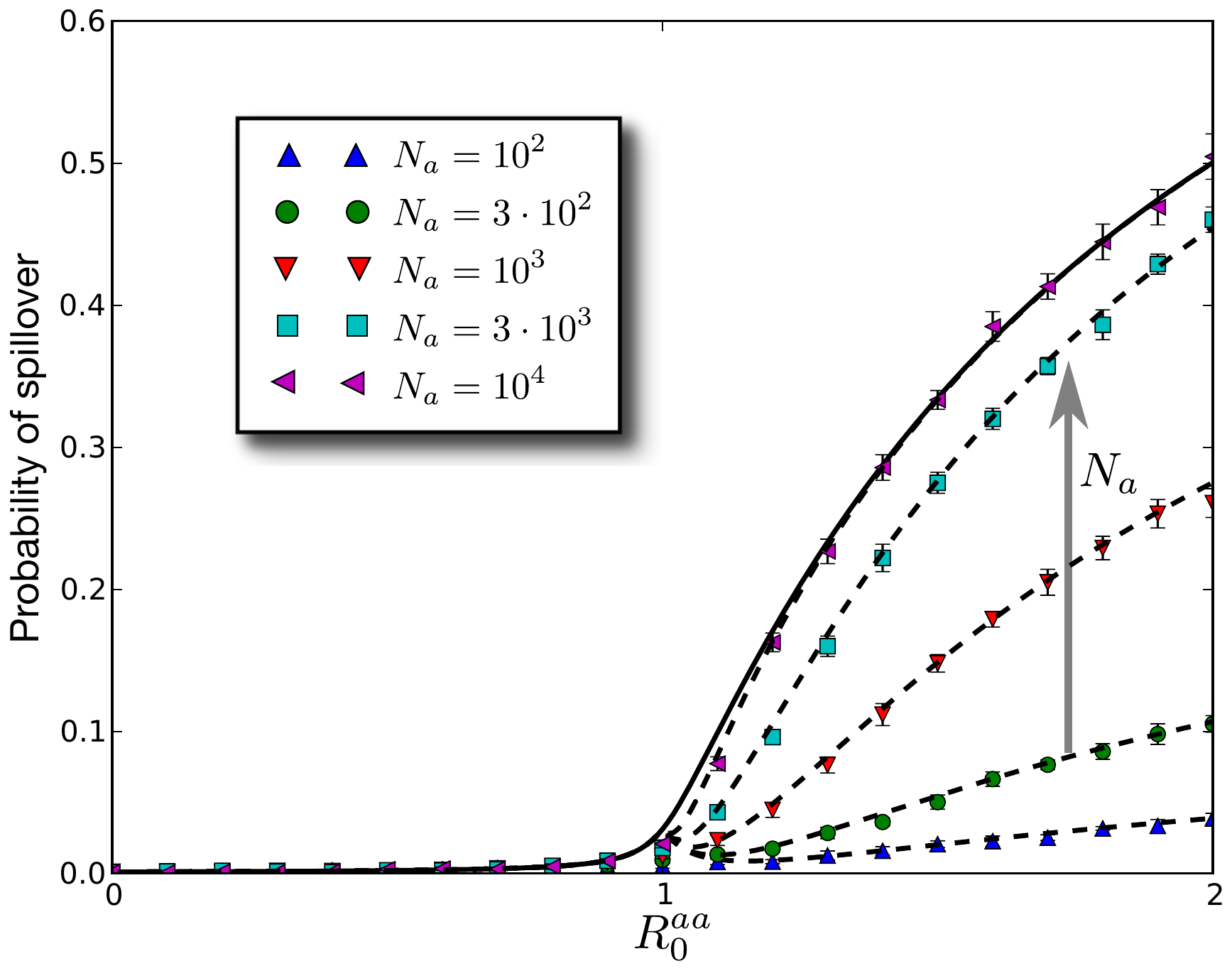}
\caption{\label{fig:Pspill_finite_size} Finite-size corrections to the probability of spillover. Dashed lines represent the analytical solution (eq. \ref{eq:Pspill_finite}) for different values of $N_a$. Solid line represents the solution from the linear birth-death process (eq. \ref{eq:P_spill}). Colored markers represents values calculated from 10,000 simulation runs done using Gillespie's direct method. All results are for fixed $R_0^{ah} = 10^{-3}$}
\end{figure}

. The probability of spillover, as calculated in eq. \ref{eq:P_spill}, is valid only in the limit of $N_a,N_h \rightarrow \infty$. Deviations from this result are expected for finite system sizes, which we report here. Using the law of total probability we can write
\begin{align} \label{eq:tot_prob}
	\mathbb{P}[\text{spill}] &= \mathbb{P}[\text{spill} \,|\, \text{small outbreak}] \!\cdot\! \mathbb{P}[\text{small outbreak}] \notag\\
	&+ \mathbb{P}[\text{spill} \,|\, \text{large outbreak}] \!\cdot\! \mathbb{P}[\text{large outbreak}]
\end{align}
where the probability is conditioned on the state of the outbreak in the animal population. Henceforth, we shall use the symbol $\mathbb{P}^{\infty}$ to represent the probability calculation done in the infinite system size limit whereas we shall use the symbol $\mathbb{P}^{N}$ for probability in the finite size calculation. 
For $R_0^{aa} \le 1$, all outbreaks are small and there are no corrections to eq. \ref{eq:P_spill}. For $R_0^{aa} > 1$, the probability of a large outbreak is non-zero. In the infinite size limit, it is implicitly assumed that $\mathbb{P}^{\infty}[\text{spill} \,|\, \text{large outbreak}] = 1$. Using this result and  $\mathbb{P}^{\infty}[\text{large outbreak}] = 1 - 1/R_0^{aa}$ in eq. \ref{eq:tot_prob}, we can calculate $\mathbb{P}^{\infty}[\text{spill} \,|\, \text{small outbreak}]$ where $R_0^{aa} > 1$.
\begin{align} \label{eq:P_spill_sm_outb}
	\mathbb{P}^{\infty}[\text{spill} \,|\, \text{small outbreak}; R_0^{aa} > 1] = 1 - R_0^{aa} v_0
\end{align}
We assume that the above result will hold for finite $N$ as well. Since small outbreaks are $o(N)$ in size, their distribution is independent of the total system size provided $N \gg 1$. More formally, we assume
\begin{equation*}
\mathbb{P}^{N}[\text{spill} \,|\, \text{small outbreak}] = \mathbb{P}^{\infty}[\text{spill} \,|\, \text{small outbreak}]
\end{equation*}
Now we calculate the finite size equivalent of $\mathbb{P}^{\infty}[\text{spill} \,|\, \text{large outbreak}]$ using the hazard function. For this calculation, we ignore the fluctuations around the mean and assume that the animal epidemic obeys the deterministic SIR. Before the first primary infection, the entire human population is susceptible and thus $S_{h,1}(t) = \nu N_h$.
\begin{align} \label{eq:p_spill_LO}
	\mathbb{P}^{N}[\text{spill} \,|\, \text{large outbreak}] &= 1 - \exp \left\{ -\int_0^{\infty} \dfrac{\beta_{ah} S_{h,1} I_a}{N_a} dt \right\} \notag\\
	&= 1 - \exp \left\{ -N_a R_0^{ah} f_a \right\}
\end{align}
where 
\begin{equation}
	f_a = \lim_{N_a \to \infty} \dfrac{\mathbb{E}[R_a(\infty)]}{N_a}
\end{equation}
is obtained by solving the final size equation for a simple SIR
\begin{equation*}
	1 - f_a = e^{-R_0 f_a}
\end{equation*}
From eq. \ref{eq:p_spill_LO}, $\mathbb{P}^{N}[\text{spill} \,|\, \text{large outbreak}] \to 1$ as $N_a \rightarrow \infty$ and this agrees with the large system size limit (eq. \ref{eq:P_spill}). Using the law of total probability, we now arrive at the probability of spillover with finite size corrections.
\begin{align} \label{eq:Pspill_finite}
	\mathbb{P}^{N}[\text{spill}; R_0^{aa} \le 1] &= 1 - v_0 \\
	\mathbb{P}^{N}[\text{spill}; R_0^{aa} > 1] &= 1 - v_0 - \left(1 - \dfrac{1}{R_0^{aa}} \right) \exp \left\{ - N_a R_0^{ah} f_a \right\} \notag
\end{align}
Figure \ref{fig:Pspill_finite_size} shows the comparison of finite size corrections as calculated using eq. \ref{eq:Pspill_finite} with stochastic simulations.

In the limit of vanishingly small $R_0^{ah}$, it is important to consider the limit of $R_0^{ah} N_a$ as $N_a \rightarrow \infty$.  Let $\xi = 
R_0^{ah} N_a$.   The probability of spillover presented in the main text (figure \ref{fig:P_spill}) assumes the limit of $\xi \rightarrow \infty$. More generally, the probability of spillover simplifies to
\begin{align}
	\lim_{\substack{R_0^{ah} \to 0 \\ N_a \to \infty}} \mathbb{P}^{N}[\text{spill}; R_0^{aa} \le 1] &= 0 \\
	\lim_{\substack{R_0^{ah} \to 0 \\ N_a \to \infty}} \mathbb{P}^{N}[\text{spill}; R_0^{aa} > 1] &= \left(1 - \dfrac{1}{R_0^{aa}} \right) \!\cdot \!\left[ 1 -  \exp \left\{ -\xi f_a \right\} \right] \notag
\end{align}
Thus, depending on the value of $\xi$, the limiting value for the probability of spillover when $R_0^{aa} > 1$ can assume any value in the range $[0, 1 - 1/R_0^{aa}]$. Thus, if $R_0^{aa} \gg 1$, then the probability of spillover is indeterminate if there is no information about $R_0^{ah} N_a$.

\subsection{Prevalence in the animal population at spillover}
\begin{figure}[tbp]
\centering
\includegraphics[width=\columnwidth]{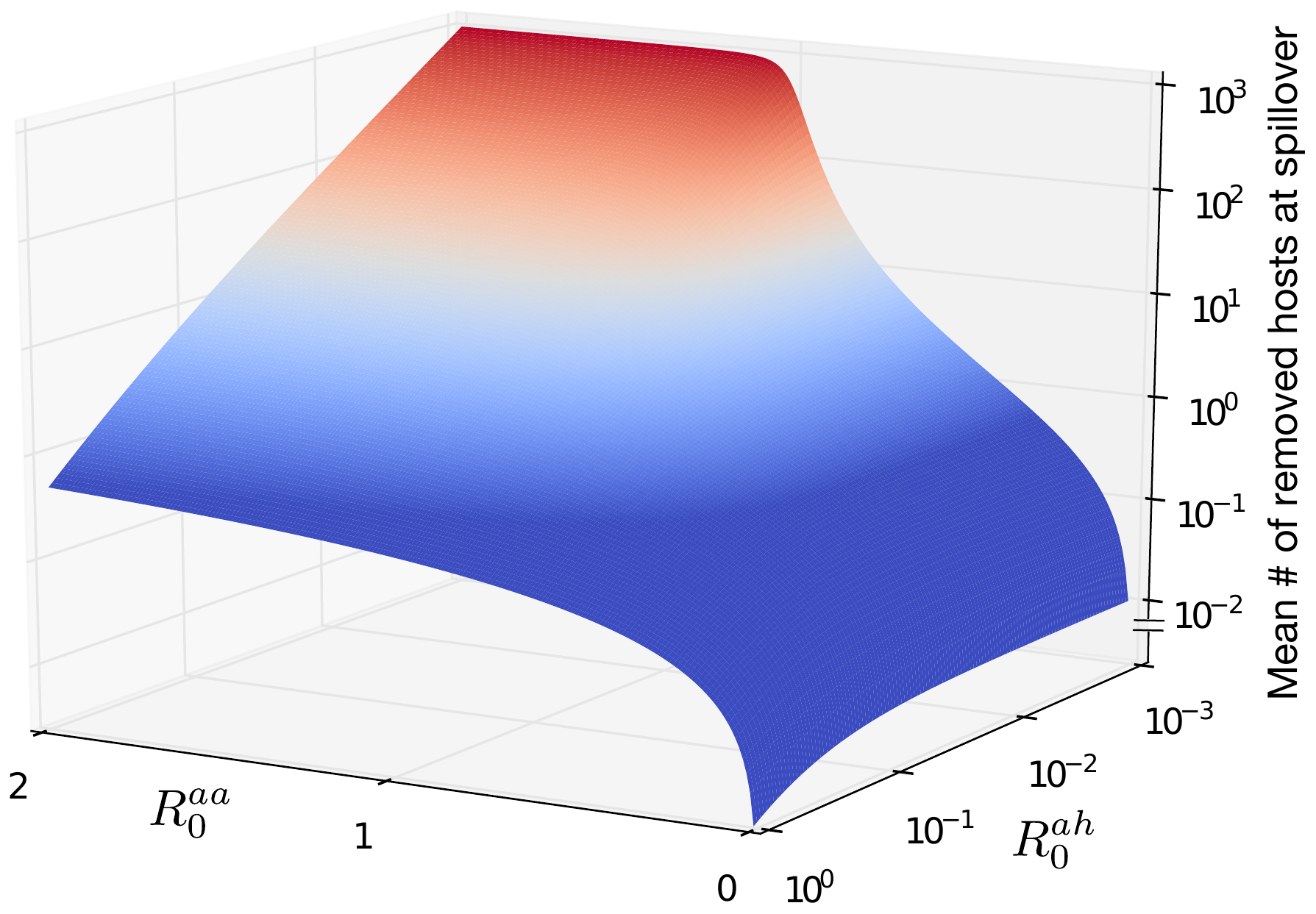}
\caption{\label{fig:mean_rec_at_spill_log} The mean number of removed animal hosts at the first passage time (obtained from eq. \ref{eq:prev_dist}) plotted as a function of $R_0^{aa}$ and $R_0^{ah}$. The surface is colored according to the standard deviation of the distribution (red:high, blue:low spanning the range $[5 \times 10^{-3},1.4 \times 10^3]$ on a $\mathrm{log}$ scale).}
\end{figure}

\begin{figure}[tbp]
\centering
\includegraphics[width=\columnwidth]{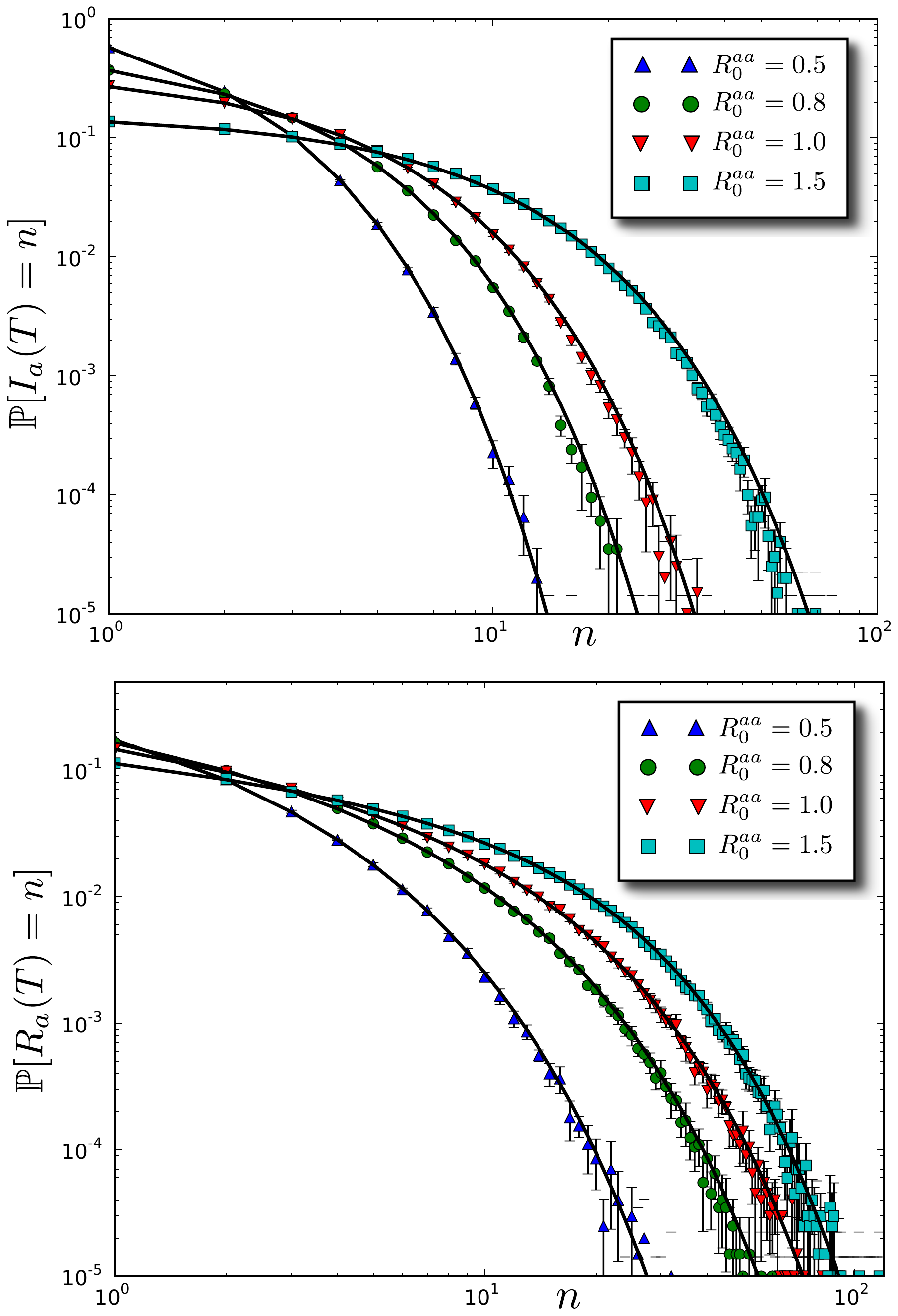}
\caption{\label{fig:prev_at_spill_sim_comp} The distribution of the number of infectious animal hosts (\emph{top}), and the number of removed animal hosts (\emph{bottom}) at first passage time $T$ for finite system size ($N_a = N_h = 1000$). Solid line represents the analytical solution obtained by sampling from the PGF in eq. \ref{eq:prev_dist}. Colored markers represents values calculated from $2 \cdot 10^5$ simulation runs done using Gillespie's direct method. All results are for fixed $R_0^{ah} = 0.1$}
\end{figure}

The distribution of infectious and removed hosts in the animal population at the first passage time can be calculated by methods outlined in \cite{SI_karlin1982linear}. By interpreting our process as linear birth-death-killing (BDK) process, the distribution of infectious hosts at spillover is the same as the distribution of killing position in the BDK process -- geometrically distributed with parameter $1 - 1/v_1$ where $v_1$ was defined in eq. \ref{eq:roots}. The calculation can be extended to include removed hosts as well (which was not part of the original results in \cite{SI_karlin1982linear}). The joint distribution of infectious and removed hosts at first passage time is generated by the following PGF.
\begin{equation} \label{eq:prev_dist}
	H_a^{S}(x,y) = \dfrac{x(v_1 - 1)}{V_1(y) - x}
\end{equation}
The surface plot for the mean number of infectious animal hosts at first passage time was shown in figure~\ref{fig:first_pass_prev_comb} (main text) and the same for the number of removed animal hosts is shown in figure~\ref{fig:mean_rec_at_spill_log}. The distribution is sampled analytically in figure~\ref{fig:prev_at_spill_sim_comp} and the results are compared with stochastic simulations for finite system sizes. As seen in the figure~\ref{fig:prev_at_spill_sim_comp} (top), the tail of the analytical distribution overestimates the prevalence slightly because of epidemic saturation that occurs in finite size SIR.

Given a prevalence of $n$ infected animal hosts at spillover (and no information about removed hosts), the maximum likelihood estimate for the parameters yields the equation $v_1 = n/(n - 1)$. From eq. \ref{eq:quad} and \ref{eq:roots}, we arrive at the following relationship between the model parameters
\begin{equation}
	R_0^{aa} = (n-1) \left(R_0^{ah} + \dfrac{1}{n} \right)
\end{equation}

\section{Branching Processes}
Here we solve the distribution of outbreak sizes for small outbreaks in the limit of large system size. An outbreak is small (or self-limited) if its size is a vanishingly small fraction of the system size in the limit $N \rightarrow \infty$. On the other hand, an outbreak whose size is a non-trivial fraction of the system size is defined a large outbreak \cite{SI_newman2002spread,SI_kenah2007}.

\subsection{Distribution of outbreak sizes} \label{sec:branching_process}
A generating function always describes the distribution of finite sized components. We shall therefore assume that the outbreaks are self-limited in this calculation. For the animal population, let $H_a(z)$ be the PGF for the distribution of outbreak sizes. From equation (\ref{eq:G_ah}), we obtain
\begin{align} \label{eq:H_a}
	H_a(z) &= G_{ah}(1,z,1;\infty) = A_a(z,1) \notag\\
			   &= \dfrac{R_0^{aa} + 1 - \sqrt{(R_0^{aa} + 1)^2 - 4 R_0^{aa} z}}{2 R_0^{aa}}
\end{align}

Let $H_{h,p}(x)$ be the PGF for the distribution of primary infections in the human population. Then, from equation (\ref{eq:G_ah}), we obtain
\begin{align} \label{eq:Hp}
	&H_{h,p}(x) = G_{ah}(1,1,x;\infty) = A_a(1,x) \\
				 &= \dfrac{R_0^{aa} \!+\! 1 \!+\! R_0^{ah}(1 \!-\! x) \!-\! \sqrt{(R_0^{aa} \!+\! 1 \!+\! R_0^{ah}(1 \!-\! x))^2 \!-\! 4 R_0^{aa} }}{2 R_0^{aa}} \notag
\end{align}

Each primary infected host in the human population acts as the progenitor for a branching process comprising of secondary infections. Let $\hat{H}_{h,s}(x)$ be the PGF for the distribution of secondary infections emanating from a primary progenitor. Then, from equation (\ref{eq:G_h})
\begin{align} \label{eq:H_hs_hat}
	\hat{H}_{h,s}(z) &= G_h(1,z;\infty) = A_h(z) \notag\\
			   &= \dfrac{R_0^{hh} + 1 - \sqrt{(R_0^{hh} + 1)^2 - 4 R_0^{hh} z}}{2 R_0^{hh} z}
\end{align}
The PGF for the joint distribution of primary and secondary infections can be written as
\begin{align} \label{eq:H_h_joint}
	H_h(x,z) = H_{h,p}(x \hat{H}_{h,s}(z))
\end{align} 
The PGF for the total number (irrespective of whether the infection was primary or secondary) is given by
\begin{align} \label{eq:H_h_tot}
	H_{h}(z) = H_{h,p}(z H_{h,s}(z))
\end{align}
Lastly, the PGF for secondary infections is given by
\begin{align} 
	H_{h,s}(z) = H_{h}(1,z)
\end{align}
Following \cite{SI_newman2001random}, we can extract probability of $n$ human hosts getting infected using Cauchy integral formula
\begin{align} \label{eq:Cauchy}
	\mathbb{P}[Z_h(\infty) = n] = \dfrac{1}{2 \pi i} \oint \dfrac{H_h(z)}{z^{n+1}} \, dz
\end{align}
where the integral is done over the unit circle $|z| = 1$ in the complex plane. Similarly, the joint probability distribution can be extracted by extending the Cauchy integral formula to higher dimensions.
\begin{align} \label{eq:cauchy_2D}
	\mathbb{P}[Z_{h,p}(\infty) \!&=\! m, Z_{h,s}(\infty) \!=\! n] = \!\dfrac{1}{(2 \pi i)^2} \!\oint \!\oint \dfrac{H_h(x,z)}{x^{m+1} z^{n+1}} \,dx\, dz
\end{align}
where the integrals are over two unit circles in the $x$ and $z$ complex planes. 

\subsection{Critical threshold}
The critical threshold is defined as the point in parameter space where the average outbreak size diverges \cite{SI_newman2002spread,SI_kenah2007} and the probability of a large outbreak becomes greater than 0. For the animal population,
\begin{align} \label{eq:avg_animal}
	\mathbb{E}[R_a(\infty)] &= H_a^{\prime}(1) \notag\\
							&= \dfrac{1}{1 - R_0^{aa}}
\end{align}
which yields the condition $R_0^{aa} = 1$ as the critical threshold. For the human population,
\begin{align} \label{eq:avg_human}
	\mathbb{E}[Z_h(\infty)] &= H_{h}^{\prime}(1) \notag\\
							&= H_{h,p}^{\prime}(1) \left\{ 1 + H_{h,s}^{\prime}(1) \right\} \notag\\
							&= \dfrac{R_0^{ah}}{(1 - R_0^{aa})(1 - R_0^{hh})}
\end{align}
From the above expression, the critical threshold for the human population is given by $\max(R_0^{aa},R_0^{hh}) = 1$. 

\subsection{Asymptotic scaling near the critical threshold}
The scaling of the outbreak sizes near the critical threshold can be investigated through the singularity analysis of the associated generated function $H(z)$ \cite{SI_flajolet2009analytic}. The dominant singularity $\zeta$ of the PGF determines the asymptotic form for $P(n)$ which is the probability of having an outbreak of size $n$. If a given PGF can be expanded around the singularity such that
\begin{equation} \label{eq:H_asym}
	H(z) \sim \left(1 - \dfrac{z}{\zeta} \right)^{\alpha}
\end{equation}
then
\begin{equation} \label{eq:Pn_asym}
	P(n) \sim  \dfrac{\zeta^{-n} n^{-\alpha - 1}}{\Gamma(-\alpha)} \;, n \to \infty
\end{equation}
where $\alpha \notin \mathbb{Z}_{> 0}$. The asymptotic form for $P(n)$ can be derived by substituting eq. \ref{eq:H_asym} in the Cauchy integral formula (eq. \ref{eq:Cauchy}) and making the following substitution
\begin{equation} \label{eq:substi}
	z \mapsto \zeta \left(1 + \dfrac{t}{n}\right)
\end{equation}
Thus, the singularity determines the exponential factor and the asymptotic form of the generating function determines the power-law exponent. By rescaling the function $H(z) \rightarrow H( z \zeta)$, the calculation of the power-law exponent is simplified since the singularity is now located at $z = 1$. We now apply this analysis to the generating function $H_h(z)$.

Let $\Delta_a = 1 - R_0^{aa}$ and $\Delta_h = 1 - R_0^{hh}$ be the distances from the critical thresholds. We first calculate the scaling near the threshold $R_0^{aa} = 1$, i.e., $\lvert \Delta_a \rvert < \lvert \Delta_h \rvert$ and $\lvert \Delta_a \rvert \ll 1$. We assume that the parameters are such that the singularities of the generating function $H_h(z)$ are far apart. The dominant singularity near the chosen threshold is given by
\begin{align} \label{eq:zeta_a}
	\zeta_a &= \left(1 + \dfrac{\left(\sqrt{R_0^{aa}} - 1\right)^2}{R_0^{ah}} \right) \left( 1 - R_0^{hh} \dfrac{\left(\sqrt{R_0^{aa}} - 1\right)^2}{R_0^{ah}} \right) \notag\\
			&= 1 + \Delta_h \left(\dfrac{\Delta_a^2}{4 R_0^{ah}} + \mathcal{O}(\Delta_a^3) \right)
\end{align}
The singularity $\zeta_a$ determines the exponential prefactor. To obtain the power-law scaling, the generating function can be analyzed at the critical point ($R_0^{aa} = 1$ in this case) without loss of generality. At the critical point $\zeta_a = 1$ and the PGF $H_{h,p}(x)$ simplifies as follows
\begin{equation}
	H_{h,p}(z) = \dfrac{2 + R_0^{ah} (1 - z) - \sqrt{R_0^{ah} (1 - z) ( 4 + R_0^{ah} (1 - z))}}{2}
\end{equation}
For further simplification, let $z\hat{H}_{h,s}(z)$ be denoted by $\tilde{H}_{h,s}(z)$. Making the substitution \ref{eq:substi} and performing a series expansion in fractional powers of $(-t/n)$ gives
\begin{equation}
	\tilde{H}_{h,s}(1 + t/n) \sim 1 + \dfrac{t}{\Delta_hn}
\end{equation}
Using \ref{eq:H_h_tot}, we obtain
\begin{align} \label{eq:H_h_exp}
	H_h(1 + t/n) \sim 1 &+ \dfrac{R_0^{ah}}{2 \Delta_h} \!\!\left(\dfrac{-t}{n} \right) - \sqrt{\dfrac{R_0^{ah}}{\Delta_h}} \!\!\left(\dfrac{-t}{n}\right)^{1/2} \!\! \notag\\
	&- \dfrac{1}{8} \left(\dfrac{R_0^{ah}}{\Delta_h}\right)^{3/2} \!\!\!\!\left(\dfrac{-t}{n}\right)^{3/2}
\end{align}
By using the Cauchy integral formula on the asymptotic expansion of $H_h(z)$, we obtain
\begin{equation}
	P_a^c(n) \sim n^{-3/2}
\end{equation}
at the threshold boundary $R_0^{aa} = 1, R_0^{hh} \neq 1$. Using the exponential prefactor obtained in eq. \ref{eq:zeta_a} we arrive at the asymptotic scaling for large $n$ near $R_0^{aa} = 1$.
\begin{equation} \label{eq:P_a}
	P_a(n) \sim \zeta_a^{-n} n^{-3/2}
\end{equation}
Note that the scaling can be guessed by looking at the leading term in the expansion, which in eq. \ref{eq:H_h_exp} is $(-t/n)^{1/2}$. Similarly, performing the same steps of analysis near the critical point of $R_0^{hh} = 1$, we obtain
\begin{equation} \label{eq:P_h}
	P_h(n) \sim \zeta_h^{-n} n^{-3/2}
\end{equation}
where 
\begin{equation*}
	\zeta_h = 1 + \dfrac{\Delta_h^2}{4}
\end{equation*}
Near the multicritical point $R_0^{aa} = R_0^{hh} = 1$, the function has a unique singularity if the value of the function $\tilde{H}_{h,s}(z)$ at its singularity $\zeta_h$ coincides with the singularity of the function $H_{h,p}(z)$, i.e.,
\begin{equation}
	1 + \dfrac{\left(\sqrt{R_0^{aa}} - 1\right)^2}{R_0^{ah}} = \dfrac{R_0^{hh} + 1}{2 R_0^{hh}}
\end{equation}
which simplifies to 
\begin{equation} \label{eq:multi_dist}
	\Delta_h = \dfrac{\Delta_a^2}{2 R_0^{ah}} + \mathcal{O}(\Delta_a^3)
\end{equation}	
for $\Delta_a, \Delta_h \ll 1$. The unique singularity is given by $\zeta_h$. Thus, the correction to the pure power-law would be $\zeta_h^{-n}$, but only on the curve given by eq. \ref{eq:multi_dist}. Next, we extract the power-law scaling at the threshold. For $R_0^{aa} = R_0^{hh} = 1$,
\begin{align}
	\tilde{H}_{h,s}(z) &= 1 - \sqrt{1 - z} \\
	H_{h,p}(z) &= \dfrac{2 + R_0^{ah} (1 - z) - \sqrt{R_0^{ah} (1 - z) ( 4 + R_0^{ah} (1 - z))}}{2} \notag
\end{align}
whose functional composition yields
\begin{align}
	H_h(z) &= H_{h,p}(\tilde{H}_{h,s}(z)) \notag\\
	H_h(z) &= \dfrac{2 + R_0^{ah} \sqrt{1 - z} - \sqrt{R_0^{ah} \sqrt{1 - z} ( 4 + R_0^{ah} \sqrt{1 - z})}}{2} \notag
\end{align}
Substituting \ref{eq:substi} and performing a series expansion in fractional powers $(-t/n)$, we obtain the $(-t/n)^{1/4}$ as the leading term. Using Cauchy integral formula, the asymptotic scaling is given by
\begin{equation} \label{eq:P_ah}
	P_{ah}^c(n) \sim n^{-5/4}
\end{equation}
Away from the multicritical threshold but staying on the curve \ref{eq:multi_dist}, the asymptotic form is
\begin{equation}
	P_{ah}(n) \sim \zeta_h^{-n} n^{-5/4}
\end{equation}
The problem of estimating the corrections to the power-law scaling away from the multi-critical point and away from the curve \ref{eq:multi_dist} is currently being investigated. In this case, the generating function will have two singularities which are coalescing at the multi-critical point. In such a scenario, there will be a crossover regime where the power-law exponent will switch from $3/2$ to $5/4$ depending on the distance from the threshold boundary.

\subsection{Finite size scaling at critical threshold}
Using the heuristic arguments presented in \cite{SI_antal2012outbreak}, we can calculate how the average outbreak size scales with system size at the threshold boundary $\max(R_0^{aa},R_0^{hh}) = 1$. For brevity, we shall adopt the following notation in this section, similar to that used in \cite{SI_antal2012outbreak}
\begin{align}
\langle n \rangle_a &\equiv \mathbb{E}[R_a(\infty)] \notag\\
\langle n \rangle_h &\equiv \mathbb{E}[R_h(\infty)]
\end{align}

\begin{figure}[tbp]
\centering
\includegraphics[width=\columnwidth]{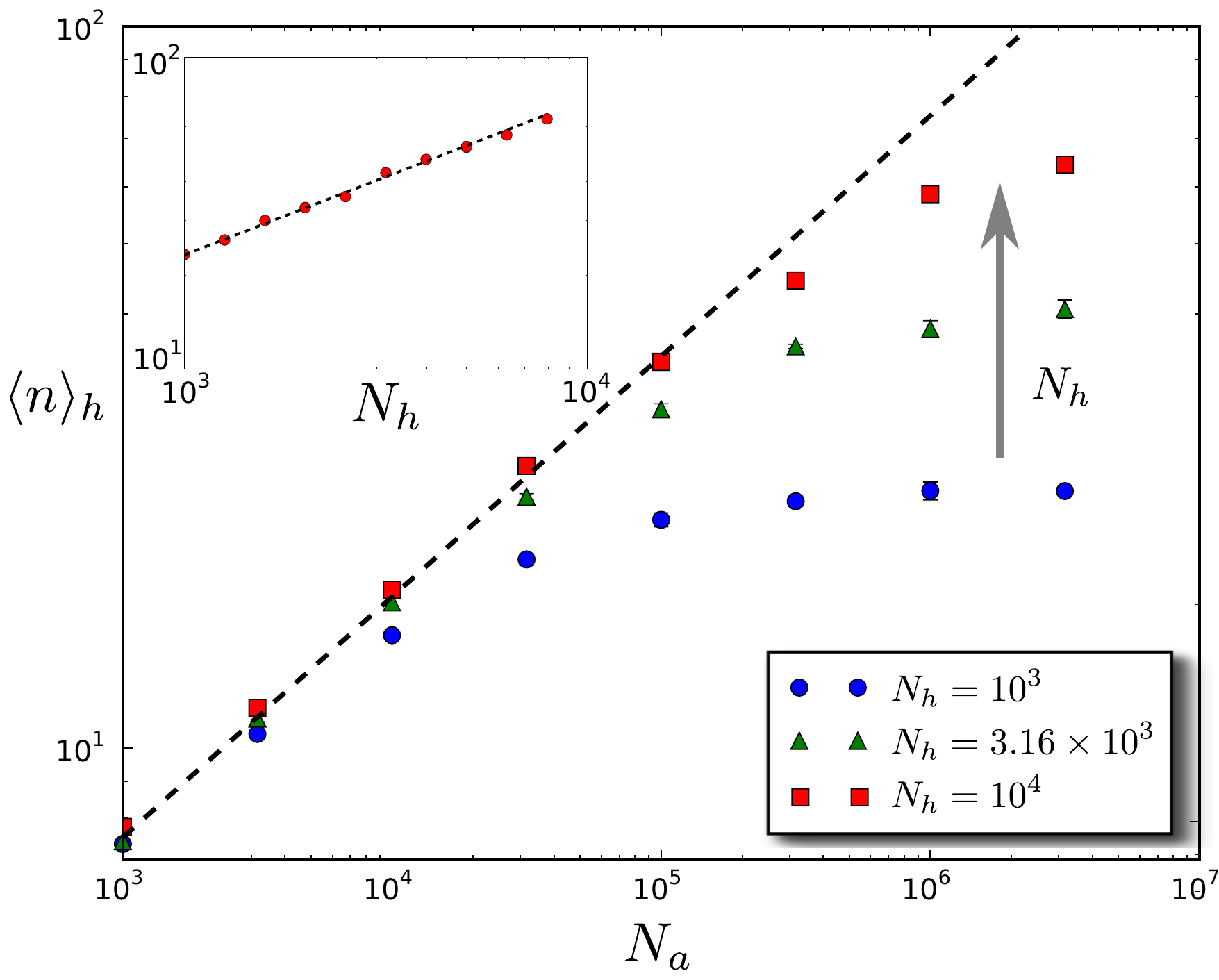}
\caption{\label{fig:finite_size_scaling} Finite size scaling at the threshold boundary $R_0^{aa} = 1, R_0^{hh} < 1$. The plot shows the scaling law for average outbreak size in humans $\langle n \rangle_h \sim N_a^{1/3}$ and crossover to $N_h^{1/2}$ when $N_h \sim N_a^{2/3}$ on a log-log plot. The points are the average of $7 \times 10^4$ stochastic realizations. The dashed line has slope 1/3. (\textit{Inset}) The average outbreak size $\langle n \rangle_h$ plotted against $N_h$ on a log-log scale for fixed $N_a = 10^7$. The dashed line has slope of 1/2. The points are the average over $10^5$ stochastic realizations. All results for $R_0^{aa} = 1, R_0^{ah} = 0.5, R_0^{hh} = 0.1$.}
\end{figure}

Let $M_a$ be the `maximal' size of an outbreak in the animal population, when $R_0^{aa} = 1$, such that an outbreak cannot exceed this size due to depletion of susceptible hosts \cite{SI_antal2012outbreak}. The effective $R_0^{aa}$ for a finite sized system reduces to
\begin{equation}
	\hat{R}_0^{aa} = 1 - M_a/N_a	
\end{equation}
Using eq. \ref{eq:avg_animal}, we obtain the following estimate for the scale of the average outbreak size
\begin{equation}
	\langle n \rangle_a \sim N_a/M_a
\end{equation}
From the 3/2 scaling law for single-type SIR \cite{SI_antal2012outbreak}, we obtain a second estimate for the average outbreak size
\begin{equation}
	\langle n \rangle_a = \sum \limits_{n=1}^{M_a} n \cdot n^{-3/2} \sim \sqrt{M_a}
\end{equation}
Equating the two estimates and imposing self-consistency, one obtains the following scaling laws (see \cite{SI_antal2012outbreak})
\begin{equation} \label{eq:scal_animal}
	M_a \sim N_a^{2/3}, \quad \langle n \rangle_a \sim N_a^{1/3}
\end{equation}
The calculation for human outbreaks is separated into 3 cases (as highlighted in figure \ref{fig:stut_composite} B,C,D). For $R_0^{aa} = 1, R_0^{hh} < 1$, the average outbreak size is given by substituting $\hat{R}_0^{aa}$ in eq. \ref{eq:avg_human}
\begin{equation} \label{eq:avg_est1}
	\langle n \rangle_h \sim N_a/M_a = N_a^{1/3}
\end{equation}
The second estimate is obtained by using the scaling law of 3/2 derived in eq. \ref{eq:P_a}.
\begin{equation} \label{eq:avg_est2}
	\langle n \rangle_h = \sum \limits_{n=1}^{M_h} n^{-1/2} \sim \sqrt{M_h}
\end{equation}
Equating the two estimates reveals $M_h \sim N_a^{2/3}$. If $\mathcal{O}(N_a) \gg \mathcal{O}(N_h^{3/2})$, the scaling relation leads to the maximal outbreak exceeding the system size, which is physically inconsistent. Thus, the maximal outbreak scale needs to be capped at $N_h$, i.e.,
\begin{equation} \label{eq:M_h2}
	M_h \sim \min(N_a^{2/3},N_h)
\end{equation}
From \ref{eq:M_h2}, we can estimate that the crossover regime between the two scales in the $\min$ function is given by $N_h \sim N_a^{2/3}$. The scaling of average outbreak size is given by $\sqrt{M_h}$, i.e.,
\begin{equation}
	\langle n \rangle_h \sim \min(N_a^{1/3},N_h^{1/2})
\end{equation}
The results are validated in figure~\ref{fig:finite_size_scaling}.
The case of $R_0^{aa} < 1, R_0^{hh} = 1$ results in the same calculations as for a single-type SIR. Thus, the scaling laws are the same as in eq. \ref{eq:scal_animal}.
\begin{equation}
	M_h \sim N_h^{2/3}, \quad \langle n \rangle_h \sim N_h^{1/3}
\end{equation}
At the multicritical point, the effective basic reproduction numbers are
\[ \hat{R}_0^{aa} = 1 - M_a/N_a, \quad \hat{R}_0^{hh} = 1 - M_h/N_h\]
From \ref{eq:avg_human}, we arrive at the first estimate
\begin{equation}
	\langle n \rangle_h \sim \dfrac{N_a}{M_a} \dfrac{N_h}{M_h} = \dfrac{N_a^{1/3} N_h}{M_h}
\end{equation}
The second estimate is derived from eq. \ref{eq:P_ah}.
\begin{equation} \label{eq:avg_est3}
	\langle n \rangle_h = \sum \limits_{n=1}^{M_h} n^{-1/4} \sim M_h^{3/4}
\end{equation}
Equating the two estimates provides the scaling for the maximal outbreak size
\begin{equation}
	M_h \sim \left( N_a^{1/3} N_h \right)^{4/7}
\end{equation}
Since the maximal outbreak can not exceed the system size
\begin{equation}
	M_h \sim \min \left(N_h, \left(N_a^{1/3} N_h \right)^{4/7} \right)
\end{equation}
The scale of the average outbreak size is given by
\begin{equation}
	\langle n \rangle_h \sim \min\left(N_h^{3/4}, \left(N_a N_h^3\right)^{1/7} \right)
\end{equation}
The crossover region in the multicritical case is $N_h \sim N_a^{4/9}$.

\subsection{Probability of large outbreak}
For the animal population, the probability of large outbreak is calculated as
\begin{align}
	\mathbb{P}[R_a(\infty) = \infty] &= 1 - H_a(1) \notag\\
							 &= 1 - \dfrac{1}{R_0^{aa}}
\end{align}
Let the probability of large human outbreak be represented by $Q$. Assuming $R_0^{ah} > 0$,
\begin{align} \label{eq:P_epd}
	Q &= 1 - H_h(1,1) \notag\\
	 &= 1 - H_{h,p}(1,H_{h,s}(1)) \\
	 &= \begin{cases}
	 		0 & \text{if $R_0^{hh} \le 1$ and $R_0^{aa} \le 1$,} \\
			1 - \dfrac{1}{R_0^{aa}} & \text{if $R_0^{hh} \le 1$ and $R_0^{aa} > 1$,} \\
			1 - A_a \left(1,\dfrac{1}{R_0^{hh}} \right) & \text{if $R_0^{hh} > 1.$} \\
		\end{cases} \notag
\end{align}

\begin{figure}[tbp]
\centering
\includegraphics[width=\columnwidth]{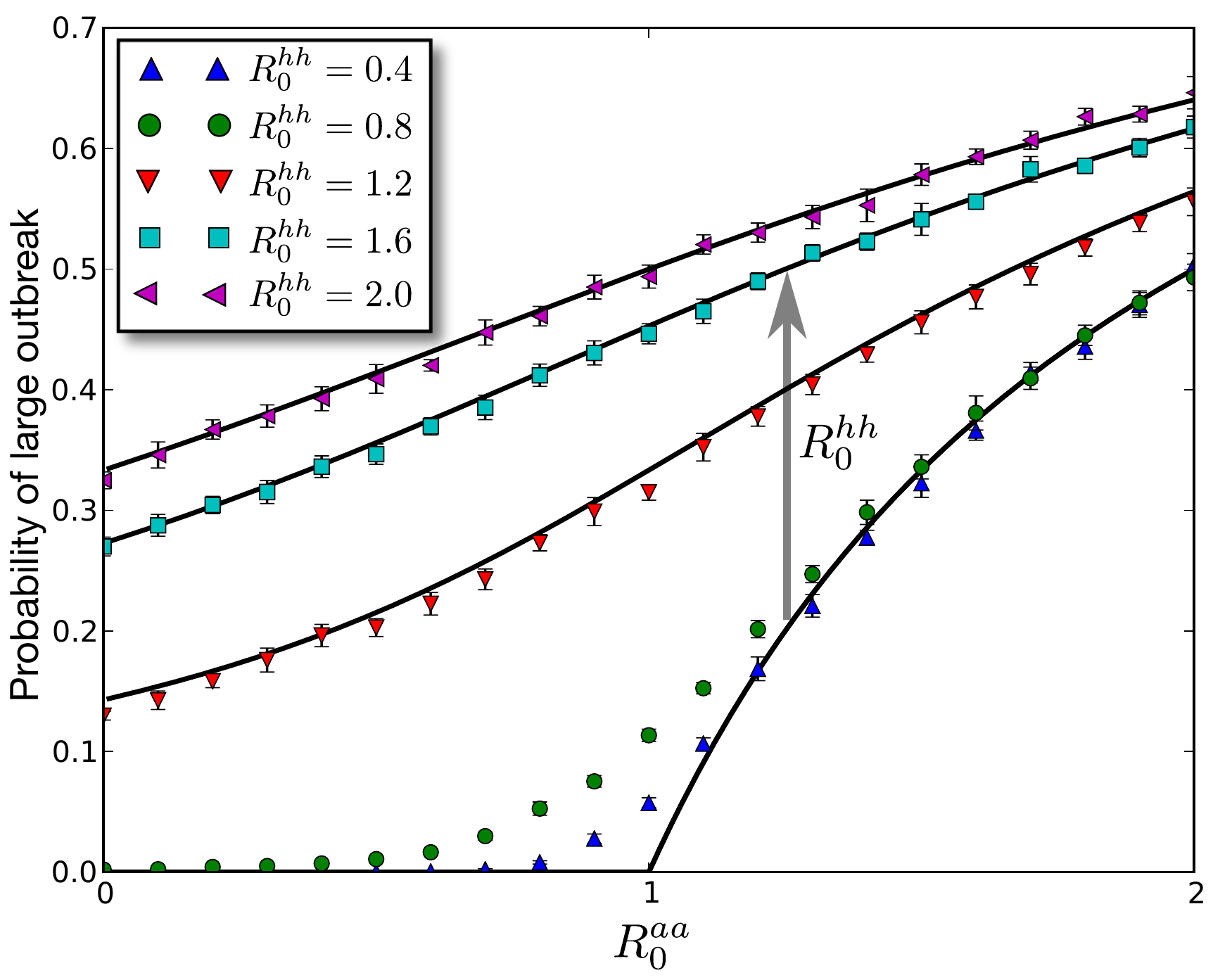}
\vspace{-0.1in}
\caption{\label{fig:Pepd_finite_size} The probability of a large human outbreak for finite populations ($N_a = N_h = 10^3$). The criteria for a large outbreak was chosen as 100 or more infected human hosts. The points represent the result of 10,000 stochastic simulations. The solid lines represent the analytical solution from eq. \ref{eq:P_epd}. The simulations do not agree with the analytical solution near the phase transition because of the chosen criteria for large outbreaks and finite size effects. All results are for fixed $R_0^{ah} =1$.}
\end{figure}

\begin{figure}[tbp]
\centering
\includegraphics[width=\columnwidth]{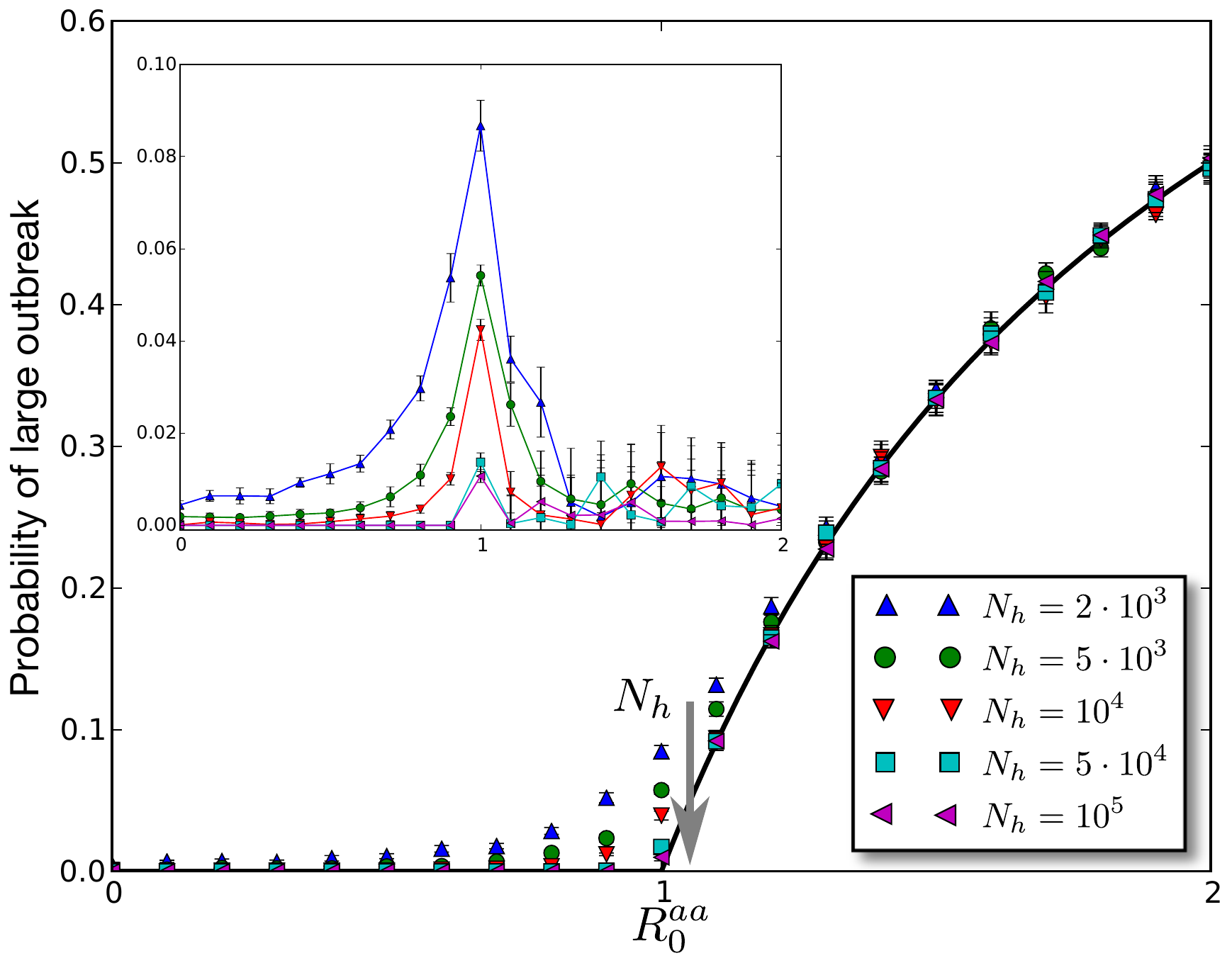}
\vspace{-0.1in}
\caption{\label{fig:Pepd_finite_size_varying_N} The probability of a large human outbreak for $R_0^{hh} = 0.8$ and varying $N_h$. The criteria for a large outbreak was chosen as the number of infected hosts being greater than 1\% of the total population. The points represent the result of 10,000 stochastic simulations. The solid line is the analytical solution $\max(0,1 - 1/R_0^{aa})$. (\emph{Inset}) The absolute difference between the analytical solution and finite size resuls. All results are for fixed $R_0^{ah} = 0.1$.}
\end{figure}

If $R_0^{hh} \le 1$, an outbreak in the human population can be large iff the outbreak in the animal population is large. In such a case, $Q$ is equal to the probability of a large outbreak in the animal population, which is a function of only $R_0^{aa}$ (see fig. \ref{fig:Pepd_finite_size_varying_N}). On the other hand, if $R_0^{hh} > 1$, a large human outbreak can occur even if the animal outbreak is small. Figures \ref{fig:Pepd_finite_size} and \ref{fig:Pepd_finite_size_varying_N} compare the analytical results with results from stochastic simulation. Away from the phase transitions at  $R_0^{aa} = 1$ and $R_0^{hh} = 1$, the results from simulation shows good agreement with the theory. Near the phase transition, the simulation results would converge to the theory for increasing $N$. Since the definition of a large outbreak becomes precise only in the limit of large system size, there are no finite size corrections that can be derived in this case.

\section{Large outbreaks}
The size of a large outbreak scales with the system size in the large population limit. The fraction of infected hosts can be calculated in several ways: (1) analytically solving the equivalent deterministic system, (2) hazard function \cite{SI_brauer2008mathematical} and (3) bond percolation on a complete graph \cite{SI_newman2002spread, SI_kenah2007network}. We use the hazard function to obtain the solution. First we write down the deterministic equations for our model.

\subsection{Deterministic Equations} \label{sec:SIR_determ}
The deterministic representation of the model can be summarized through the following system of ODEs.
\begin{align} \label{eq:SIR_determ}
	&\dfrac{d\mathcal{S}_a}{d\tau} = -R_0^{aa} \mathcal{S}_a \mathcal{I}_a \notag\\
	&\dfrac{d \mathcal{I}_a}{d\tau} = R_0^{aa} \mathcal{S}_a \mathcal{I}_a - \mathcal{I}_a \notag\\
	&\dfrac{d \mathcal{R}_a}{d\tau} = \mathcal{I}_a \notag\\ \notag\\
	&\dfrac{d \mathcal{S}_{h,1}}{d\tau} \!=\! - \lambda \, \mathcal{S}_{h,1} \mathcal{I}_a \!-\! \kappa \, R_0^{hh} \mathcal{S}_{h,1} (\mathcal{I}_{h,1,p} \!\!+\! \mathcal{I}_{h,1,s} \!\!+\! \mathcal{I}_{h,2}) \notag\\
	&\dfrac{d \mathcal{I}_{h,1,p}}{d\tau} = \lambda \, \mathcal{S}_{h,1} \mathcal{I}_a - \kappa \, \mathcal{I}_{h,1,p} \notag\\
	&\dfrac{d \mathcal{I}_{h,1,s}}{d\tau} = \kappa \, R_0^{hh}   \mathcal{S}_{h,1} (\mathcal{I}_{h,1,p} + \mathcal{I}_{1,s} + \mathcal{I}_{h,2}) - \kappa \, \mathcal{I}_{h,1,s} \\
	&\dfrac{d \mathcal{R}_{h,1,p}}{d\tau} = \kappa \,\mathcal{I}_{h,1,p} \notag\\
	&\dfrac{d \mathcal{R}_{h,1,s}}{d\tau} = \kappa \,\mathcal{I}_{h,1,s} \notag\\\notag\\
	&\dfrac{d \mathcal{S}_{h,2}}{d\tau} = - \kappa \, R_0^{hh} \mathcal{S}_{h,2} (\mathcal{I}_{h,1,p} + \mathcal{I}_{1,s} + \mathcal{I}_{h,2}) \notag\\
	&\dfrac{d \mathcal{I}_{h,2}}{d\tau} =  \kappa \, R_0^{hh} \mathcal{S}_{h,2} (\mathcal{I}_{h,1,p} + \mathcal{I}_{1,s} + \mathcal{I}_{h,2}) - \kappa \, \mathcal{I}_{h,2} \notag\\
	&\dfrac{d \mathcal{R}_{h,2}}{d\tau} = \kappa \, \mathcal{I}_{h,2} \notag
\end{align}
where the variables $\mathcal{S}_{\star},\mathcal{I}_{\star},\mathcal{R}_{\star}$ are non-dimensional state variables that have been normalized by the total population of the species. Here, all dynamical variables for the human population are normalized by $N_h$ and time is normalized by the average infectious period of the animal hosts. Two new variables are introduced here
\begin{equation}
	\lambda = \dfrac{\rho R_0^{ah}}{\nu}, \quad \kappa = \dfrac{\gamma_h}{\gamma_a}
\end{equation}
The non-dimensional parameters governing the dynamics of the system are: $(R_0^{aa},\lambda, R_0^{hh}, \kappa)$. The initial conditions that we use to solve this system are given below 
\[ \mathcal{S}_a(0) = 1 - \dfrac{1}{N_a},\; \mathcal{I}_a(0) = \dfrac{1}{N_a}, \; \mathcal{R}_a(0) = 0\]
\[ \mathcal{S}_{h,1}(0) = \nu,\; \mathcal{S}_{h,2}(0) = 1 - \nu, \; \mathcal{I}_{h,\star}(0) = 0, \; \mathcal{R}_{h,\star}(0) = 0\]

\subsection{Mean final size}
Let $f_{\star}$ be the relative size of the infected hosts in the various host compartments in the limit of large system size for the stochastic version of the model.
\begin{align}
	f_a &= \lim_{N_a \rightarrow \infty} \dfrac{\mathbb{E}[R_{a}(\infty)]}{N_a} \notag\\
	f_{h,p} &= \lim_{N_h \rightarrow \infty} \dfrac{\mathbb{E}[R_{h,1,p}(\infty)]}{N_h} \notag\\
	f_{h,s} &= \lim_{N_h \rightarrow \infty} \dfrac{\mathbb{E}[R_{h,1,s}(\infty)] + \mathbb{E}[R_{h,2}(\infty)]}{N_h} \\
	f_{h,1} &= \lim_{N_h \rightarrow \infty} \dfrac{\mathbb{E}[R_{h,1,p}(\infty)] + \mathbb{E}[R_{h,1,s}(\infty)]}{N_h} \notag\\
	f_{h,2} &= \lim_{N_h \rightarrow \infty} \dfrac{\mathbb{E}[R_{h,2}(\infty)]}{N_h} \notag\\
	f_h &= f_{h,1} + f_{h,2} \notag\\ \notag\\[-10pt]
		&= f_{h,p} + f_{h,s} \notag
\end{align}
Using survival analysis described in \cite{SI_brauer2008mathematical}, we 
proceed with calculations for the various $f_\star$. The calculation is based on the result that in the limit of large system size the final epidemic size is the same as that given by solving the deterministic system of equations, i.e.,
\begin{equation}
	f_{\star} = \mathcal{R}_\star(\infty)
\end{equation}
For a randomly chosen susceptible host in the animal population, the cumulative hazard function is the probability of not getting infected before time $t$. This function can be calculated as follows
\begin{align} 
	\Lambda_{aa}(t) = e^{- \int_0^t \beta_{aa} \mathcal{I}_a ds}
\end{align}
At steady state, the hazard function simplifies as follows
\begin{equation} \label{eq:lambda_aa}
	\Lambda_{aa}(\infty) = e^{- R_0^{aa} \mathcal{R}_a(\infty)}	= e^{-R_0^{aa} f_a} 
\end{equation} 
The probability of escaping infection would be $1 - f_a$. Equating this with equation (\ref{eq:lambda_aa}), we obtain
\begin{align} \label{eq:f_a}
	1 - f_a &= e^{-R_0^{aa} f_a}
\end{align}
Similarly for the human hosts, we first look at type 1 hosts (who are at risk of both primary and secondary transmissions). The hazard functions for the animal to human and human to human transmissions are given by
\begin{align}
	\Lambda_{ah}(\infty) &= e^{-\lambda f_a} \notag\\
	\Lambda_{hh}(\infty) &= e^{-R_0^{hh} f_h}
\end{align}
A randomly chosen type 1 human host will not be infected during a large outbreak only if it escapes getting infected from both the primary and secondary transmissions.
\begin{equation} \label{eq:f_h1}
f_{h,1} = \nu \left( 1 - e^{-\lambda f_a} e^{-R_0^{hh} f_h} \right)
\end{equation}
The prefactor $\nu$ is to normalize the relative size of the epidemic by size of the population of type 1 human hosts. Similarly, we can calculate the size of the epidemic in type 2 hosts.
\begin{align} \label{eq:f_h2}
	f_{h,2} = (1 - \nu) \left(1 - e^{-R_0^{hh} f_h} \right)
\end{align}
The total size of the epidemic in the human population is obtained by adding equations \ref{eq:f_h1} and \ref{eq:f_h2}
\begin{align} \label{eq:f_h}
	f_h &= f_{h,1} + f_{h,2} \notag\\
	f_h &= 1 - \left( 1 - \nu + \nu e^{-\lambda f_a} \right) e^{-R_0^{hh} f_h}
\end{align}
The solution of the implicit equation \ref{eq:f_a} feeds in to equation \ref{eq:f_h} whose solution can then be used to solve equations \ref{eq:f_h1} and \ref{eq:f_h2}. In the absence of secondary transmissions, i.e, $R_0^{hh} = 0$, the epidemic in the type 1 hosts would only consist of primary infections. Let this fraction of infected hosts be denoted by $f^0_{h,p}$, which can be obtained by setting $R_0^{hh}$ to 0 in equation \ref{eq:f_h1}.
\begin{align} \label{eq:zeta_hp}
	f^0_{h,p} = \nu \left( 1 - e^{-\lambda f_a} \right)
\end{align}
Immediately comparing equations \ref{eq:f_h1} and \ref{eq:zeta_hp}, we can assert that
\begin{equation} \label{eq:zeta_hp_lt_f_h1}
	f^0_{h,p} \le f_{h,1}
\end{equation}
with the equality holding for $R_0^{hh} = 0$. Note that $f_{h,p} \neq f^0_{h,p}$ since $f^0_{h,p}$ is the size of the epidemic in the absence of human to human transmissions whereas $f_{h,p}$ is the size of the epidemic when both forces of infection are active. In the latter scenario, the two forces of infection would be competing for a susceptible. Thus, the proportion of the epidemic caused by primary infections would be reduced as compared to the case where only the primary transmission is active.
\begin{equation} \label{eq:f_hp_lt_zeta_hp}
	f_{h,p} \le f^0_{h,p}
\end{equation}
For the last part of the analysis, consider a randomly chosen infected type 1 human host $i$. This host is exposed to both primary and secondary forces of infection. Let $T_{h,p}^{(i)}$ be the time when this host receives disease via a primary transmission. Similarly, let $T_{h,s}^{(i)}$ be the time when the host receives disease via a secondary transmission. If $T_{h,p}^{(i)} < T_{h,s}^{(i)}$, a primary infection is realized else a secondary infection is realized. Note that the idea of multiple transmissions is a mathematical construct rather than a biological realism. A host that has already been infected and recovered can not be infected again (in the SIR framework). But the host is still subjected to the second force of infection which can result in another successful (albeit redundant) transmission. From the analogy with reaction kinetics \cite{SI_keelingmodeling}, it is important to know which transmission reaction fired first since that would determine whether the infection was primary or secondary. We can now write down an expression for the relative size of the epidemic consisting of primary infections.
\begin{align} \label{eq:f_hp_gt_f_hps}
	&f_{h,p} = \nu \cdot \, \mathbb{P}[T_{h,p}^{(i)} < T_{h,s}^{(i)}] \\
			&= \nu \!\left( \mathbb{P}[T_{h,p}^{(i)} \!<\! T_{h,s}^{(i)},T_{h,s}^{(i)} \!=\! \infty] + \mathbb{P}[T_{h,p}^{(i)} \!<\! T_{h,s}^{(i)}, T_{h,s}^{(i)} \!<\! \infty] \right) \notag\\
			&= \nu\,e^{-R_0^{hh} f_h} \left(1 - e^{-\lambda f_a} \right) f_{h,1} + \nu\, \mathbb{P}[T_{h,p}^{(i)} < T_{h,s}^{(i)} < \infty] \notag\\
			&\ge f_{h,p \backslash s} \notag
\end{align}
where
\begin{equation} \label{eq:f_hp_not_s}
	f_{h,p \backslash s} \equiv \nu \,e^{-R_0^{hh} f_h} \left(1 - e^{-\lambda f_a} \right) f_{h,1}
\end{equation}
Combining equations \ref{eq:zeta_hp_lt_f_h1}, \ref{eq:f_hp_lt_zeta_hp} and \ref{eq:f_hp_not_s}, we get
\begin{equation} \label{eq:f_hp_ineq}
	f_{h,p \backslash s} \; \le \; f_{h,p} \; \le \; f^0_{h,p} \; \le \; f_{h,1}
\end{equation}
where the equality holds for $R_0^{hh} = 0$.

\subsection{Primary vs Secondary}
\begin{figure}[tbp]
\centering
\includegraphics[width=\columnwidth]{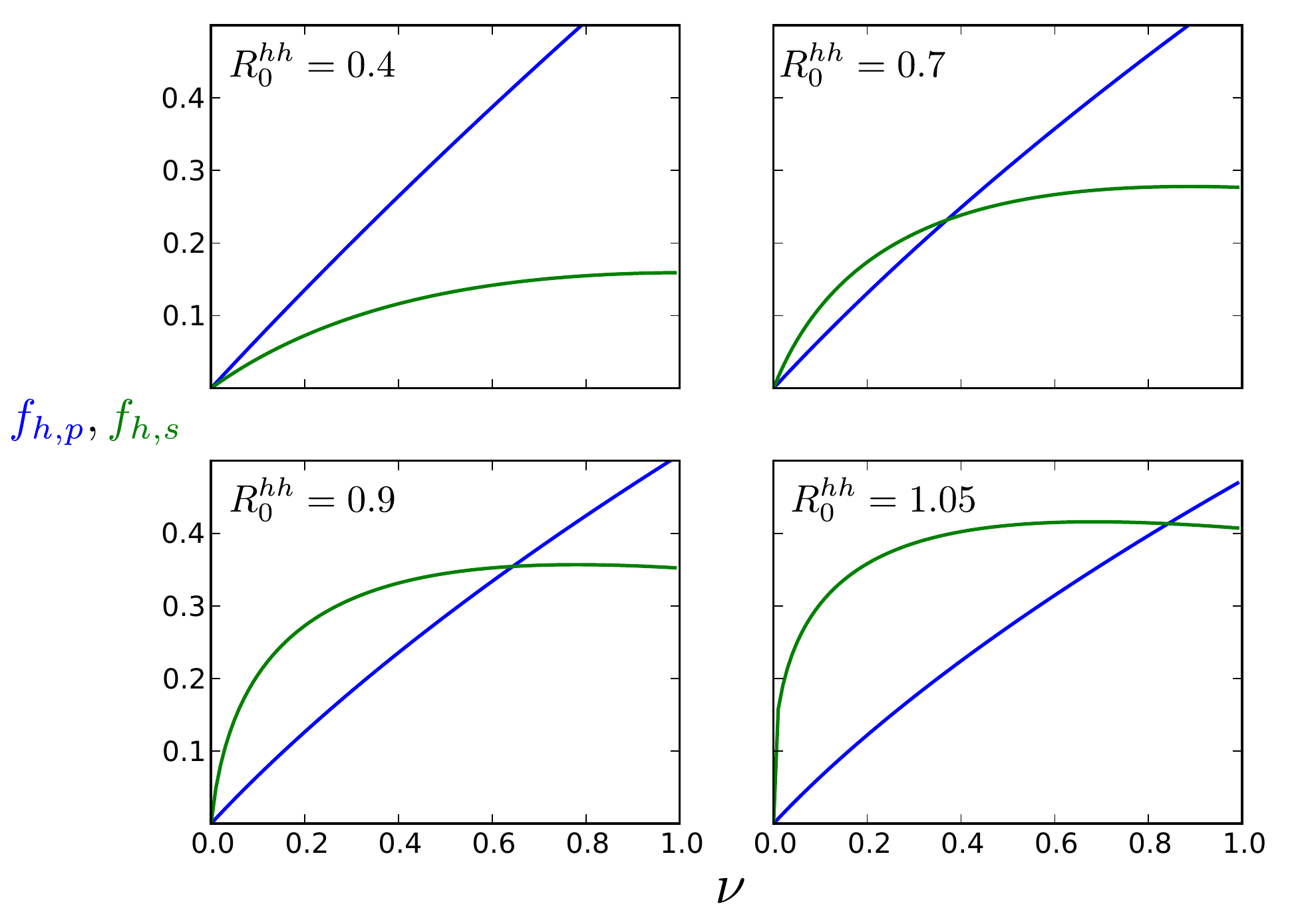}
\vspace{-0.1in}
\caption{\label{fig:prim_vs_sec} Fraction of human hosts infected
  during a large outbreak via a primary transmission ($f_{h,p}$, blue), and secondary transmission ($f_{h,s}$, green) plotted for different
  values of $\nu$ and $R_0^{hh}$. Analytical solution obtained from
  solving the deterministic system. Remaining parameters for the plots: $N_a = N_h = 1000, \beta_{aa} = 2.0, \hat{\beta}_{ah} = 1.5, \gamma_a = \gamma_h = 1.0$}
\end{figure}

Figure~\ref{fig:prim_vs_sec} shows the average number of primary and
secondary infections occurring during a large outbreak for different values of $\nu$ and $R_0^{hh}$. The solutions were obtained by solving the deterministic equations (eq. \ref{eq:SIR_determ}). The curve for the primary infections will always be non-decreasing with $\nu$. This follows from intuition that as more and more susceptible hosts become at risk, the number of primary infections will also increase. The fact that the effective $R_0$ for the A-H transmissions, i.e., $R_0^{ah}$  also increases with $\nu$ compounds the effect. The secondary infections on the other hand exhibit non-monotonicity in some regions of parameter space. This can be attributed to the love-hate relationship between the two forces of infection (\emph{ah} and \emph{hh}) acting on susceptible human hosts. On one hand, the secondary infections cannot occur unless there are primary infections. Thus, for small values of $\nu$, there is a strong correlation between the number of primary and secondary infections. On the other hand, as $\nu$ increases, the two forces start competing for the same susceptible hosts. Depending on the model parameters, either of the two forces can dominate in different regions of the phase space which leads to the rich behavior for the number of secondary infections.

\subsection{Bifurcation point}
As evident from figure~\ref{fig:prim_vs_sec}, the curves $f_{h,p}$ and $f_{h,s}$ when plotted against $\nu$ may or may not intersect apart from $\nu = 0$. Here we calculate the condition under which the bifurcation would occur creating a second point of intersection. Since we do not have an explicit expression for $f_{h,p}$ or $f_{h,s}$, the solution is not rigorous. But numerical experiments over a large parameter ranges have revealed that the solution does hold. The solution assumes that both $f_{h,p}$ and $f_{h,s}$ are concave functions of $\nu$. For small values of $R_0^{hh}$, we can assert that the secondary infections would be smaller than primary infections for all values of $\nu$. Thus $\nu = 0$ would be the only solution. As we increase $R_0^{hh}$, a bifurcation would occur at $\nu = 0$ and a second solution would emerge. At the bifurcation point, the slope of $f_{h,p}$ and $f_{h,s}$ would be equal. Thus, the bifurcation condition is
\begin{equation} \label{eq:bifur}
	\dfrac{\partial f_{h,p}}{\partial \nu} \Bigg |_{\nu = 0} = \dfrac{\partial f_{h,s}}{\partial \nu} \Bigg |_{\nu = 0}
\end{equation}
Since we don't have an analytical expression for $f_{h,p}$, we will work with equation \ref{eq:f_hp_ineq}. Assuming $R_0^{hh} < 1$, from equation \ref{eq:f_h} we get
\begin{align}
	f_h |_{\nu = 0} &= 0 \notag\\
	\dfrac{\partial f_h}{\partial \nu} \Big |_{\nu=0} &= \dfrac{1}{1 - R_0^{hh}}
\end{align}
Using the above solutions in \ref{eq:zeta_hp} and \ref{eq:f_hp_gt_f_hps}, we obtain
\begin{align} \label{eq:sandwich}
	\dfrac{\partial f_{h,p \backslash s}}{\partial \nu} \Big |_{\nu=0} = \dfrac{\partial f^0_{h,p}}{\partial \nu} \Big |_{\nu=0} = 1
\end{align}
From equation \ref{eq:sandwich} and \ref{eq:f_hp_ineq}, we obtain.
\begin{align} \label{eq:df_hp_0}
	\dfrac{\partial f_{h,p}}{\partial \nu} \Big |_{\nu=0} &= 1
\end{align}
For $f_{h,s}$,
\begin{align} \label{eq:df_hs_0}
	\dfrac{\partial f_{h,s}}{\partial \nu} \Big |_{\nu=0} &= \dfrac{\partial f_h}{\partial \nu} \Big |_{\nu=0} - \dfrac{\partial f_{h,p}}{\partial \nu} \Big |_{\nu=0} \notag\\[10pt]
	&= \dfrac{R_0^{hh}}{1 - R_0^{hh}}
\end{align}
Equating \ref{eq:df_hp_0} and \ref{eq:df_hs_0}, we obtain $R_0^{hh} = 1/2$ is the bifurcation point where the two slopes are equal. For $R_0^{hh} < 1/2$, the number of secondary transmissions will always be smaller than primary ones for $\nu > 0$. For $R_0^{hh} > 1/2$, the two curves will either intersect or $f_{h,s}$ will be strictly greater than $f_{h,p}$. We were unable to calculate analytically the point $\nu^{\star}$ of intersection of the two curves or the point in parameter space where the point of intersection disappears.

\subsection{Non-identifiability of epidemic driver}
\begin{figure}[htp]
\centering
\includegraphics[width=\columnwidth]{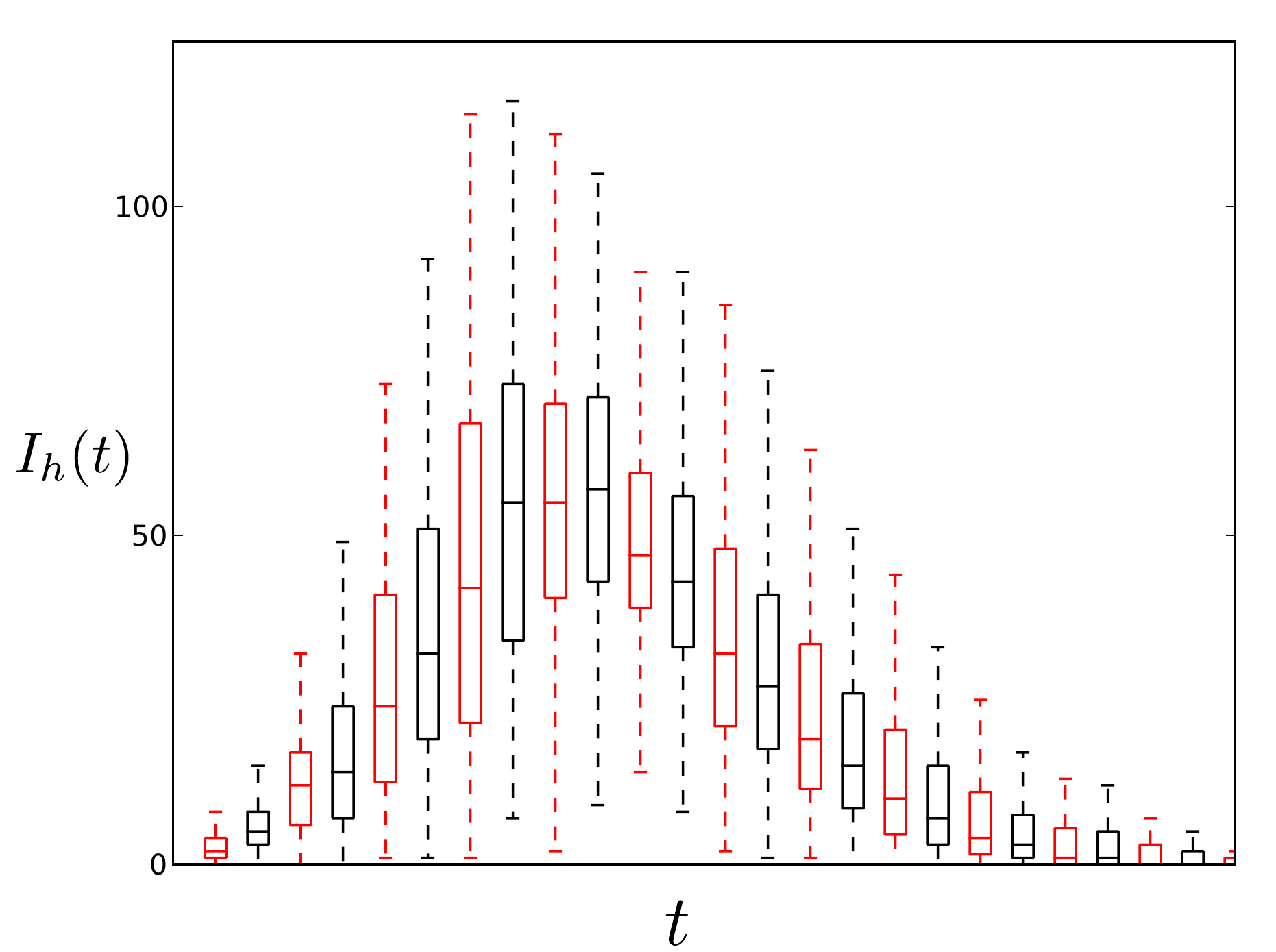}
\vspace{-0.1in}
\caption{\label{fig:sim_comp} Box plots for simulated epidemic
trajectories for the human population in each of the following two
scenarios: (black) the epidemic is driven by only A-H transmissions
with no H-H transmissions ($\beta_{aa} \!\!=\!\! \hat{\beta}_{ah} \!\!=\!\! 1.5,\, \beta_{hh} \!\!=\!\! 0$, one infected animal host at $t=0$) ; and (red) the epidemic is driven by only H-H transmissions after initial spillover ($\beta_{aa} \!=\! \hat{\beta}_{ah} = 0,\, \beta_{hh} \!=\! 1.5$, one infected human host at $t=0$). The remaining parameters for the simulations: $N_a = N_h = 1000, \nu = 1, \gamma_a = \gamma_h = 1.0$.}
\end{figure}

We present the argument in the main text that given just the time series data for a large outbreak it is not possible to identify whether the epidemic is driven by a large animal outbreak or by human to human transmission. To make our case, we compare the distribution of stochastic epidemic profiles for the mentioned scenarios in figure~\ref{fig:sim_comp}: one where there are only primary infections taking place and other where there is only human to human transmission. As can be seen in the plot, the distribution of the infection profile is almost identical for the chosen sets of parameters. The parameter combinations chosen are not necessarily unique and such non-identifiability can occur by choosing parameters from different parts of the phase space.

\subsection{Variance of the final size}
Since our model fits in the formalism of a generalized multi-type epidemic, we shall borrow notation and results from \cite{SI_ball1993final,SI_britton2002epidemics}. We have 3 host types in our system: animal, type 1 human and type 2 humans which we shall label as 1, 2 and 3 in this section. The populations for the respective types are
\[N_1 = N_a, \quad N_2 = \nu N_h, \quad N_3 = (1 - \nu) N_h\]
Let $\pi_i$ be the fraction of hosts in each type,
\[ \pi_1 = \dfrac{N_a}{N_a + N_h}, \quad \pi_2 = \dfrac{\nu N_h}{N_a + N_h}, \quad \pi_3 = \dfrac{(1-\nu) N_h}{N_a + N_h} \]
and $F_i$ be the fraction of individuals infected in each type in the limit of large (but finite) population.
\[ F_1 \!=\! F_a \!=\! \dfrac{R_a(\infty)}{N_a}, \; F_2 \!=\! \dfrac{R_{h,1}(\infty)}{\nu N_h},\]
\[F_3 \!=\! \dfrac{R_{h,2} (\infty)}{(1 \!-\! \nu) N_h}, \; F_h \!=\! \dfrac{R_h(\infty)}{N_h}\]
and $\phi_i$ be the mean fraction.
\[ \phi_1 = \phi_a = f_a, \; \phi_2  = \dfrac{f_{h,1}}{\nu}, \; \phi_3 = \dfrac{f_{h,2}}{1 - \nu}, \; \phi_h = f_h \]
Next we define $\Omega$ matrix as
\begin{equation*}
	\boldsymbol{\Omega} = 
	\begin{bmatrix}
		R_0^{aa} \left(1 + \dfrac{1}{\rho}\right) & \dfrac{R_0^{ah}}{\nu} \left(1 + \rho\right) & 0 \\ \\[-5pt]
		0 & R_0^{hh} \left(1 + \rho\right) & R_0^{hh} \left(1 + \rho\right) \\ \\[-5pt]
		0 & R_0^{hh} \left(1 + \rho\right) & R_0^{hh} \left(1 + \rho\right)
	\end{bmatrix}
\end{equation*}
A central limit theorem in Ball and Clancy \cite{SI_ball1993final} shows that the vector $\left\{\sqrt{N_j}(F_j - \phi_j), j=1,2,3 \right\}$
is asymptotically Gaussian with mean $\boldsymbol{0}$ and variance matrix
\[ \Sigma = {S^T}^{-1} \Xi S^{-1}\]
where the matrices $S$ and $\Xi$ are given by
\begin{align}
	S_{ij} &= \delta_{ij} - \sqrt{\pi_i \pi_j} \Omega_{ij} (1 - \phi_j) \\
	\Xi_{ij} &= \phi_i(1 - \phi_j) \delta_{ij} + \sqrt{\pi_i \pi_j} (1 - \phi_i) (1 - \phi_j) \!\!\sum \limits_{k=1,2,3} \pi_k \phi_k \Omega_{ki} \Omega_{kj} \notag
\end{align}
where $\delta_{ij}$ is the \emph{Kronecker} delta. We have used the fact that for the exponentially distributed infectious period such as in our case the mean is equal to the standard deviation. Therefore the term $(\sigma_k/\mu_k)$ which is present in general form of the expression given in \cite{SI_britton2002epidemics}, does not appear here. Since the disease process in the animal population can be treated as a single type epidemic process, the variance can be written explicitly.
\begin{align}
	\mathrm{Var}(F_a) = \dfrac{f_a(1 - f_a) \left[1 + (R_0^{aa})^2(1 - f_a)\right]}{N_a \left(1 - R_0^{aa}(1 - f_a) \right)^2}
\end{align}
For the human population, 
\[ F_h = \nu F_2 + (1 - \nu) F_3 \]
The variance of the relative final size in the human population can then be calculated as
\begin{align} \label{eq:f_h_var}
	\mathrm{Var}(F_h) = \dfrac{\nu \, \Xi_{22} + (1-\nu) \, \Xi_{33} + 2\sqrt{\nu(1 - \nu)} \, \Xi_{23}}{N_h}
\end{align}
The special case of $\nu = 1$ (which eliminates type 2 hosts) yields the following expression for above
\begin{align}
	\mathrm{Var}(F_h) &= \dfrac{f_h(1 - f_h) \left[1 + (R_0^{hh})^2(1 - f_h)\right]}{N_h \left(1 - R_0^{hh}(1 - f_h) \right)^2} \notag\\
					  &+ \dfrac{\rho (R_0^{ah})^2 f_a (2 - f_a) (1 - f_h)^2 }{N_h \left(1 - R_0^{aa}(1 - f_a) \right)^2 \left(1 - R_0^{hh}(1 - f_h) \right)^2}
\end{align}


\begin{thebibliography}{10}

\bibitem{woolhouse2006host}
Woolhouse, MEJ \& Gowtage-Sequeria, S
\newblock (2005) Host range and emerging and reemerging pathogens.
\newblock {\em Emerg.\ Infect.\ Dis.} {\bf 11}, 1842--1847.

\bibitem{kuiken2005pathogen}
Kuiken, T et~al.
\newblock (2005) Pathogen surveillance in animals.
\newblock {\em Science} {\bf 309}, 1680--1681.

\bibitem{lloyd2009epidemic}
Lloyd-Smith, JO et~al.
\newblock (2009) Epidemic dynamics at the human-animal interface.
\newblock {\em Science} {\bf 326}, 1362--1367.

\bibitem{greger2007human}
Greger, M
\newblock (2007) The human/animal interface: emergence and resurgence of
  zoonotic infectious diseases.
\newblock {\em Crit.\ Rev.\ Microbiol} {\bf 33}, 243--299.

\bibitem{jones2008global}
Jones, KE et~al.
\newblock (2008) Global trends in emerging infectious diseases.
\newblock {\em Nature} {\bf 451}, 990--993.

\bibitem{onehealth}
Conti, LA \& Rabinowitz, PM
\newblock (2011) One health initiative.
\newblock {\em Infektolo{\v{s}}ki Glasnik} {\bf 31}, 176--178.

\bibitem{wood2012framework}
Wood, JLN et~al.
\newblock (2012) A framework for the study of zoonotic disease emergence and
  its drivers: spillover of bat pathogens as a case study.
\newblock {\em Philos.\ Trans.\ R.\ Soc.\ Lond.\ B Biol.\ Sci} {\bf 367},
  2881--2892.

\bibitem{karesh2012ecology}
Karesh, WB et~al.
\newblock (2012) Ecology of zoonoses: natural and unnatural histories.
\newblock {\em Lancet} {\bf 380}, 1936--1945.

\bibitem{wolfe2007origins}
Wolfe, ND, Dunavan, CP,  \& Diamond, J
\newblock (2007) Origins of major human infectious diseases.
\newblock {\em Nature} {\bf 447}, 279--283.

\bibitem{morse2012prediction}
Morse, SS et~al.
\newblock (2012) Prediction and prevention of the next pandemic zoonosis.
\newblock {\em Lancet} {\bf 380}, 1956--1965.

\bibitem{allen2012mathematical}
Allen, LJS et~al.
\newblock (2012) Mathematical modeling of viral zoonoses in wildlife.
\newblock {\em Nat.\ Resour.\ Model.} {\bf 25}, 5--51.

\bibitem{collinge2006disease}
Collinge, SK \& Ray, C
\newblock (2006) {\em Disease Ecology: Community Structure and Pathogen
  Dynamics}.
\newblock (Oxford University Press, USA).

\bibitem{childs2007introduction}
Childs, JE, Richt, JA,  \& Mackenzie, JS
\newblock (2007) Introduction: conceptualizing and partitioning the emergence
  process of zoonotic viruses from wildlife to humans.
\newblock {\em Curr.\ Top.\ Microbiol.\ Immunol.} {\bf 315}, 1--31.

\bibitem{epstein1995emerging}
Epstein, PR
\newblock (1995) Emerging diseases and ecosystem instability: new threats to
  public health.
\newblock {\em Am.\ J.\ Public Health} {\bf 85}, 168--172.

\bibitem{brauer2008mathematical}
{Van den Driessche}, FBP, Wu, J,  \& Allen, LJS
\newblock (2008) {\em {Mathematical Epidemiology}}.
\newblock (Springer).

\bibitem{sethna2006statistical}
Sethna, JP
\newblock (2006) {\em Statistical mechanics: entropy, order parameters, and
  complexity}.
\newblock (Oxford Univ.\ Press).

\bibitem{antal2012outbreak}
Antal, T \& Krapivsky, PL
\newblock (2012) Outbreak size distributions in epidemics with multiple stages.
\newblock {\em J.\ Stat.\ Mech.} {\bf 2012}, P07018.

\bibitem{blumberg2013inference}
Blumberg, S \& Lloyd-Smith, JO
\newblock (2013) Inference of \uppercase{R0} and transmission heterogeneity from the size
  distribution of stuttering chains.
\newblock {\em PLoS Comput. Biol.} {\bf 9}, e1002993.

\bibitem{bailey1990elements}
Bailey, NTJ
\newblock (1990) {\em {The Elements of Stochastic Processes with Applications
  to the Natural Sciences}}.
\newblock (Wiley-Interscience).

\bibitem{griffiths1972bivariate}
Griffiths, DA
\newblock (1972) A bivariate birth-death process which approximates to the
  spread of a disease involving a vector.
\newblock {\em J.\ Appl.\ Probab.} {\bf 9}, 65--75.

\bibitem{karlin1982linear}
Karlin, S \& Tavar{\'e}, S.
\newblock (1982) Linear birth and death processes with killing.
\newblock {\em J.\ Appl.\ Probab.} {\bf 19}, 477--487.

\bibitem{athreya1972ney}
Athreya, KB \& Ney, PE
\newblock (1972) {\em {Branching Processes}}.
\newblock (Springer-Verlag Berlin Heidelberg).

\bibitem{ball1993final}
Ball, F \& Clancy, D
\newblock (1993) The final size and severity of a generalised stochastic
  multitype epidemic model.
\newblock {\em Adv.\ Appl.\ Probab.} {\bf 25}, 721--736.

\bibitem{britton2002epidemics}
Britton, T
\newblock (2002) Epidemics in heterogeneous communities: estimation of
  \uppercase{R0} and secure vaccination coverage.
\newblock {\em J.\ R.\ Stat.\ Soc.\ Series B Stat.\ Methodol.} {\bf 63},
  705--715.

\bibitem{antia2003role}
Antia, R, Regoes, RR, Koella, JC,  \& Bergstrom, CT
\newblock (2003) The role of evolution in the emergence of infectious diseases.
\newblock {\em Nature} {\bf 426}, 658--661.

\bibitem{restif2012model}
Restif, O et~al.
\newblock (2012) Model-guided fieldwork: practical guidelines for
  multidisciplinary research on wildlife ecological and epidemiological
  dynamics.
\newblock {\em Ecol.\ Lett.} {\bf 15}, 1083--1094.

\bibitem{gillespie1977exact}
Gillespie, DT
\newblock (1977) Exact stochastic simulation of coupled chemical reactions.
\newblock {\em J.\ Phys.\ Chem.} {\bf 81}, 2340--2361.

\bibitem{keelingmodeling}
Keeling, MJ \& Rohani, P
\newblock (2008) {\em {Modeling Infectious Diseases in Humans and Animals}}.
\newblock (Princeton Univ.\ Press).

\end{thebibliography}

\begin{thebibliography}{10}

\bibitem{SI_athreya1972ney}
Athreya, KB \& Ney, PE
\newblock (1972) {\em {Branching Processes}}.
\newblock (Springer-Verlag Berlin Heidelberg).

\bibitem{SI_bailey1990elements}
Bailey, NTJ
\newblock (1990) {\em {The Elements of Stochastic Processes with Applications
  to the Natural Sciences}}.
\newblock (Wiley-Interscience).

\bibitem{SI_lloyd2009epidemic}
Lloyd-Smith, JO et~al.
\newblock (2009) Epidemic dynamics at the human-animal interface.
\newblock {\em Science} {\bf 326}, 1362--1367.

\bibitem{SI_antal2011exact}
Antal, T \& Krapivsky, PL
\newblock (2011) Exact solution of a two-type branching process: models of
  tumor progression.
\newblock {\em J.\ Stat.\ Mech.} {\bf 2011}, P08018.

\bibitem{SI_griffiths1972bivariate}
Griffiths, DA
\newblock (1972) A bivariate birth-death process which approximates to the
  spread of a disease involving a vector.
\newblock {\em J.\ Appl.\ Probab.} {\bf 9}, 65--75.

\bibitem{SI_karlin1982linear}
Karlin, S \& Tavar{\'e}, S
\newblock (1982) Linear birth and death processes with killing.
\newblock {\em J.\ Appl.\ Probab.} {\bf 19}, 477--487.

\bibitem{SI_keelingmodeling}
Keeling, MJ \& Rohani, P
\newblock (2008) {\em {Modeling Infectious Diseases in Humans and Animals}}.
\newblock (Princeton Univ.\ Press).

\bibitem{SI_newman2002spread}
Newman, MEJ
\newblock (2002) {Spread of epidemic disease on networks}.
\newblock {\em Phys.\ Rev.\ E Stat.\ Nonlin.\ Soft Matter Phys.} {\bf 66},
  16128.

\bibitem{SI_kenah2007}
Kenah, E \& Robins, JM
\newblock (2007) Second look at the spread of epidemics on networks.
\newblock {\em Phys.\ Rev.\ E Stat.\ Nonlin.\ Soft Matter Phys.} {\bf 76},
  036113.

\bibitem{SI_newman2001random}
Newman, MEJ, Strogatz, SH,  \& Watts., DJ
\newblock (2001) Random graphs with arbitrary degree distributions and their
  applications.
\newblock {\em Phys.\ Rev.\ E Stat.\ Nonlin.\ Soft Matter Phys.} {\bf 64},
  026118.

\bibitem{SI_flajolet2009analytic}
Flajolet, P \& Sedgewick, R
\newblock (2009) {\em Analytic combinatorics}.
\newblock (Cambridge Univ.\ Press).

\bibitem{SI_antal2012outbreak}
Antal, T \& Krapivsky, PL
\newblock (2012) Outbreak size distributions in epidemics with multiple stages.
\newblock {\em J.\ Stat.\ Mech.} {\bf 2012}, P07018.

\bibitem{SI_brauer2008mathematical}
{Van den Driessche}, FBP, Wu, J,  \& Allen, LJS
\newblock (2008) {\em {Mathematical Epidemiology}}.
\newblock (Springer).

\bibitem{SI_kenah2007network}
Kenah, E \& Robins, JM
\newblock (2007) Network-based analysis of stochastic {SIR} epidemic models with random and proportionate mixing.
\newblock {\em J.\ Theor.\ Biol.} {\bf 249}, 706.

\bibitem{SI_ball1993final}
Ball, F \& Clancy, D
\newblock (1993) The final size and severity of a generalised stochastic
  multitype epidemic model.
\newblock {\em Adv.\ Appl.\ Probab.} {\bf 25}, 721--736.

\bibitem{SI_britton2002epidemics}
Britton, T
\newblock (2002) Epidemics in heterogeneous communities: estimation of
  \uppercase{R0} and secure vaccination coverage.
\newblock {\em J.\ R.\ Stat.\ Soc.\ Series B Stat.\ Methodol.} {\bf 63},
  705--715.

\end{thebibliography}
\end{document}